\documentclass[conference]{IEEEtran}
\IEEEoverridecommandlockouts
\usepackage{verbatim}
\usepackage{graphicx}
\usepackage{epstopdf}
\usepackage{balance}
\usepackage{comment}
\usepackage{subfigure}
\usepackage{booktabs}
\usepackage{amsmath}
\usepackage{amssymb}
\usepackage{bm}
\usepackage{colortbl}
\usepackage{tabularx}
\usepackage{color}
\usepackage{epsfig}
\usepackage{xspace}
\usepackage{placeins}
\usepackage{multirow}
\usepackage{enumitem}
\usepackage{leading}
\def\BibTeX{{\rm B\kern-.05em{\sc i\kern-.025em b}\kern-.08em
    T\kern-.1667em\lower.7ex\hbox{E}\kern-.125emX}}
		
		\makeatletter
  \newcommand\figcaption{\def\@captype{figure}\caption}
  \newcommand\tabcaption{\def\@captype{table}\caption}
\makeatother

\begin{document}

\title{Measurement, Characterization and Modeling of LoRa Technology in Multi-floor Buildings}

\author{\IEEEauthorblockN{Weitao Xu$^{12}$, Jun Young Kim$^{2}$, Walter Huang$^{2}$, Salil Kanhere$^{1}$, Sanjay Jha$^{1}$, Wen Hu$^{12}$}
\IEEEauthorblockA{$^{1}$University of New South Wales, Australia \\ $^{2}$WBS Technology Australia\\
		\{firstname.lastname\}@unsw.edu.au, \{jun.kim,walter\}@wbstech.com.au  \\
}
}
\vspace{-0.2in}
\maketitle

\begin{abstract}
In recent years, we have witnessed the rapid development of LoRa technology, together with extensive studies trying to understand its performance in various application settings. In contrast to measurements performed in large outdoor areas, limited number of attempts have been made to understand the characterization and performance of LoRa technology in indoor environments. In this paper, we present a comprehensive study of LoRa technology in multi-floor buildings. Specifically, we investigate the large-scale fading characteristic, temporal fading characteristic, coverage and energy consumption of LoRa technology in four different types of buildings. Moreover, we find that the energy consumption using different parameter settings can vary up to 145 times. These results indicate the importance of parameter selection and enabling LoRa adaptive data rate feature in energy-limited applications. We hope the results in this paper can help both academia and industry understand the performance of LoRa technology in multi-floor buildings to facilitate developing practical indoor applications.
\end{abstract}


\section{Introduction}
\label{sec:introduction}
The Internet of Things (IoT) brings the promise of a world comprising smart cities, smart buildings, smart homes to improve every aspect of our lifestyle. The rapid development of various IoT applications has created the requirement of new wireless technologies that can provide cost effective large area coverage. Low Power Wide Area Networks (LPWANs) communication technologies have recently emerged as a viable alternative to cellular and mesh networks to fulfill the vast requirements. LPWANs are designed to fill the gap between short-range, high-bandwidth networks (e.g, Bluetooth, WiFi, and ZigBee) and cellular networks (e.g., GSM and LTE)~\cite{blenn2017lorawan}. 
Many LPWANs applications are proving its cost efficiency and large-scale IoT application suitability. Examples of such LPWAN technologies include LoRa~\cite{lora}, Sigfox~\cite{Sigfox} and NB-IoT~\cite{NBiot}. Among these competing LPWAN technologies, LoRa is attracting attention primarily since it offers affordable connectivity to the low-power devices distributed over large geographical areas. 

Despite its young age, LoRa has grown rapidly and drawn wide attention in the past few years. Extensive research has already been carried out in outdoor environment to understand the performance of LoRa technology~\cite{petajajarvi2015coverage,kartakis2016demystifying,gaelens2017lora,augustin2016study}. It is reported that the communication range can be up to 15Km with over $60\%$ delivery rate in Line-of-Sight (LOS) conditions in rural areas~\cite{petajajarvi2015coverage}. Apart from the applications in wide areas, such as smart city and smart traffic, the LoRa technology can also become the enablers for new applications in indoor environment such as gas meter monitoring, home automation and smart buildings. However, according to our survey, the performance evaluation of LoRa in indoor environment either use limited number of nodes~\cite{petajajarvi2015coverage} or focus on only few metrics such as Received Signal Strength Indicator (RSSI) and Signal Noise Ratio (SNR)~\cite{blenn2017lorawan}. Few attempts have been made to comprehensively study the propagation, characterization and performance of LoRa technology in multi-floor buildings, especially buildings with multi-story basements.
 
Our industry partner WBS Technology\footnote{https://wbstech.com.au/} is a smart building solution provider in Australia. We have implemented and deployed a LoRa-based smart building network in 9 production smart building of different types. It is known that different buildings have different communication conditions because of the difference in size, shape and structure. In order to make sure the designed LoRa network works at its optimal communication mode, it is crucial to understand the characteristics and performance of LoRa in different types of buildings. To this end, we perform a detailed study to understand the propagation, characteristics and performance of LoRa technology in different multi-floor buildings. In particular, we deployed a LoRa testbed which consists of 10 LoRa nodes in one of the testing buildings. To the best of our knowledge, this is the first attempt to comprehensively investigate the large-scale fading, temporal fading, coverage and energy consumption of LoRa technology in different multi-floor buildings. With the help of results in this paper, the LoRa network deployed in 9 different buildings has successfully run for over 6 months (for more details please visit the homepage of WBS Technology). We also hope that these results can provide insights for other companies who are interested in using LoRa in their applications. 

The remainder of this paper is organised as follows. Sec.~\ref{config} describes the experimental setup and Sec.~\ref{sec:method} presents the measurement methodology. Followed by that, the results of large-scale fading characterization, temporal fading characterization, and coverage experiment are presented in Sec.~\ref{sec:largescaleresults},~\ref{sec:temporalfadingresults} and~\ref{sec:coverageresults}, respectively. Finally, Sec.~\ref{related} discusses the related work and Sec.~\ref{sec:conclusion} concludes the paper.
\begin{figure*}[!t]
	\centering
	\subfigure[LoRa mote]{
		\includegraphics[width=1.3in,height=1.1in]{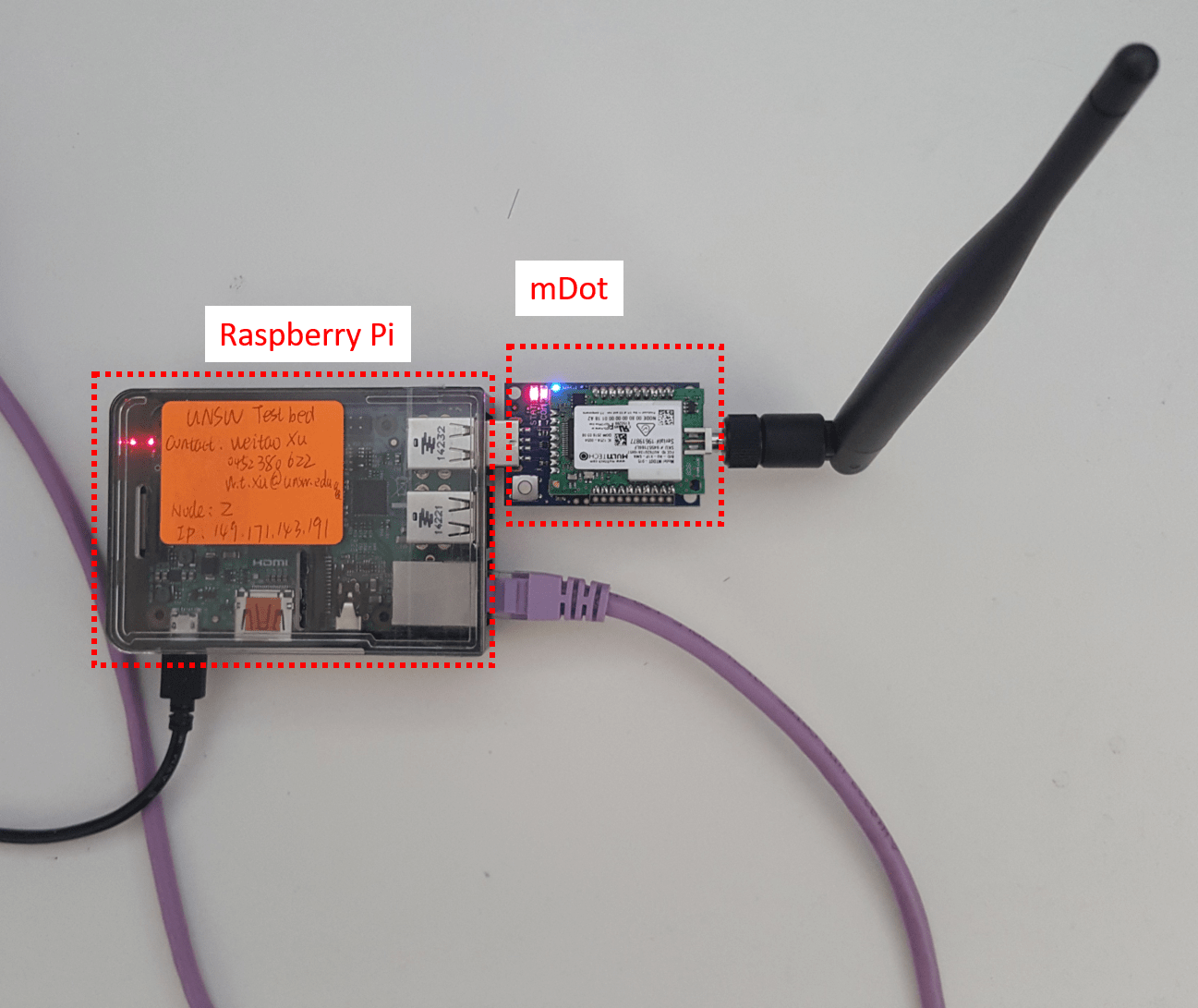}
		\label{fig:hardware}}
		\hspace{-0.1in}
	\subfigure[Office Building]{
		\includegraphics[width=1.3in,height=1.1in]{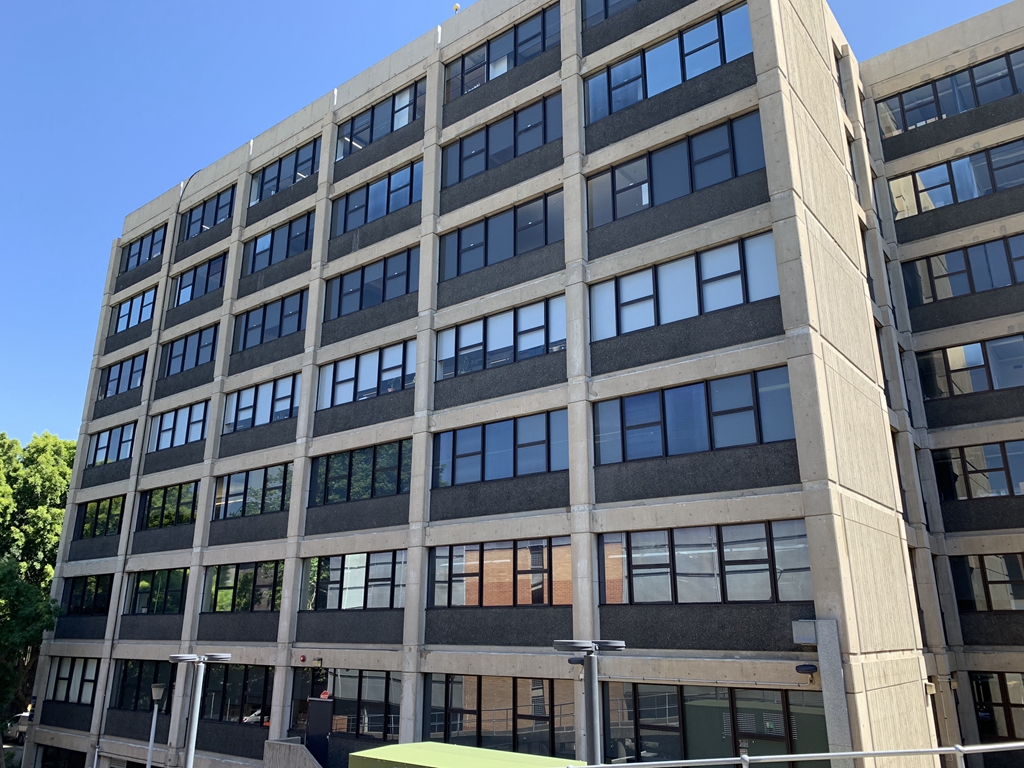}
		\label{fig:officebuilding}}
		\hspace{-0.1in}
		\subfigure[Residential Building]{
		\includegraphics[width=1.3in,height=1.1in]{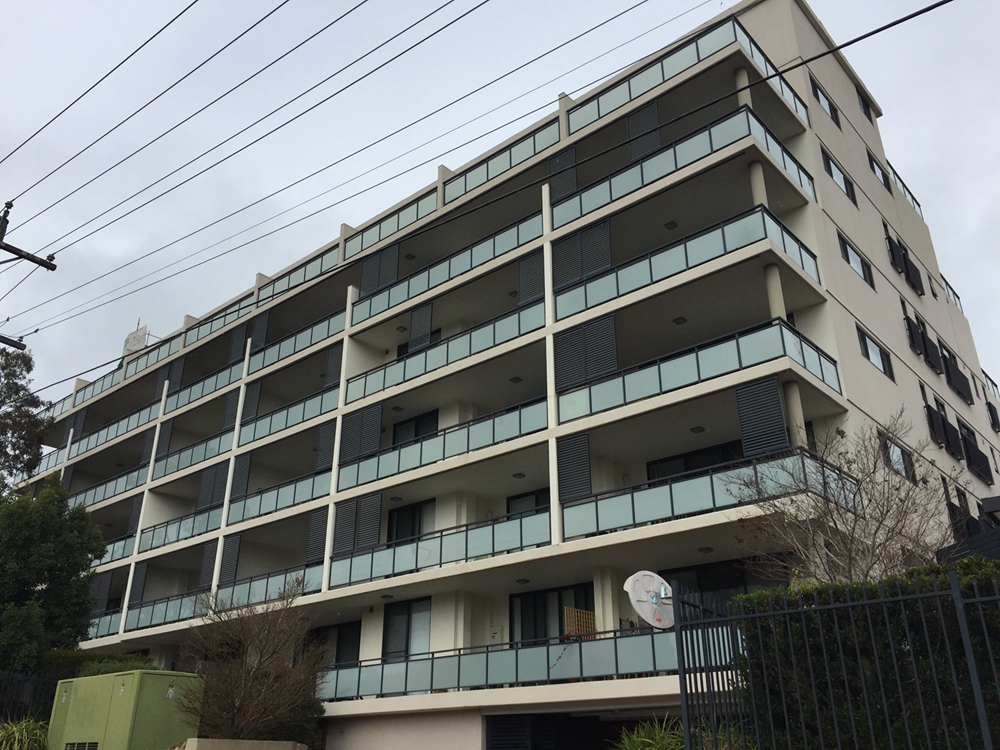}
		\label{fig:residentialbuilding}}
		\hspace{-0.1in}
        \subfigure[Car park]{
		\includegraphics[width=1.3in,height=1.1in]{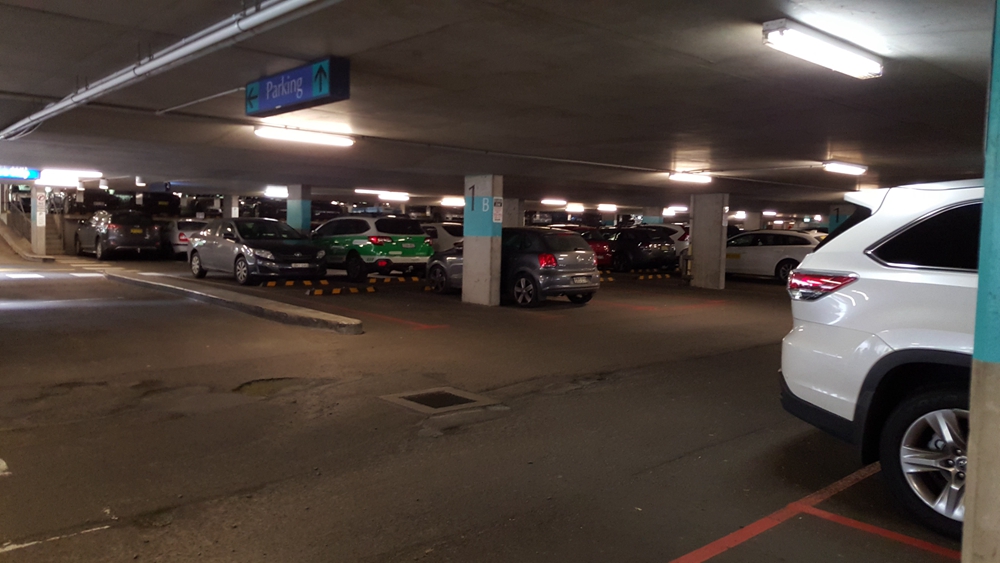}
		\label{fig:unswcarpark}}
		\hspace{-0.1in}
		\subfigure[Warehouse]{
		\includegraphics[width=1.3in,height=1.1in]{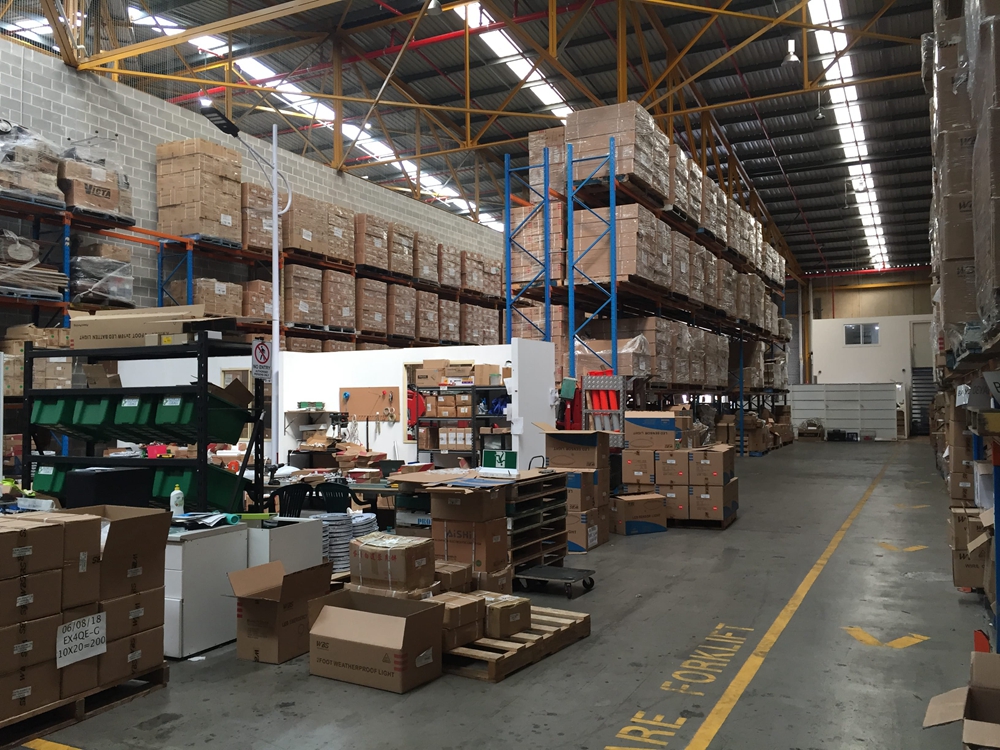}
		\label{fig:warehouse}}
	\subfigure[Floor plan of Level 4]{
		\includegraphics[width=2.5in]{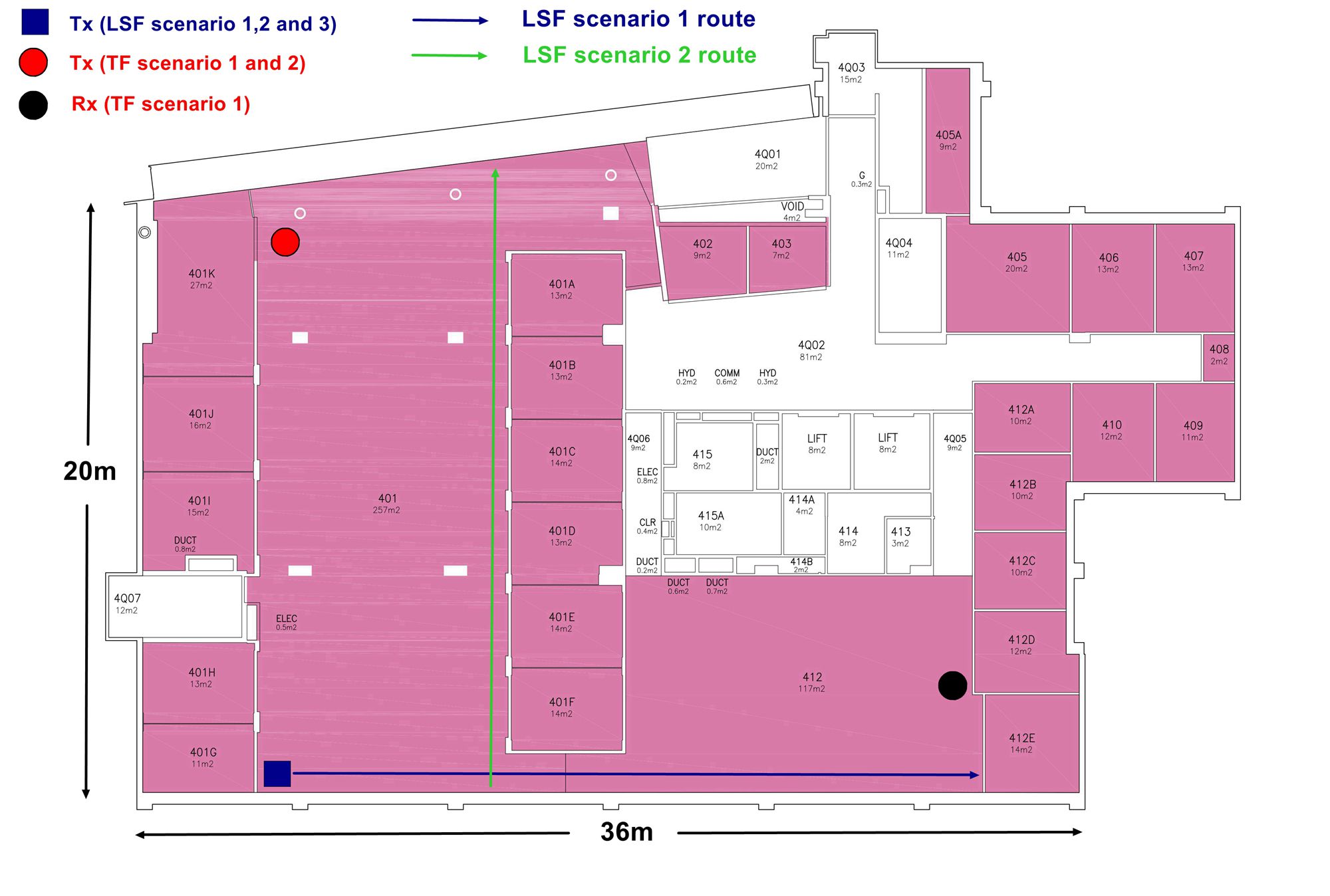}
		\label{fig:level4}}
	\subfigure[Floor plan of ground level]{
		\includegraphics[width=2.5in]{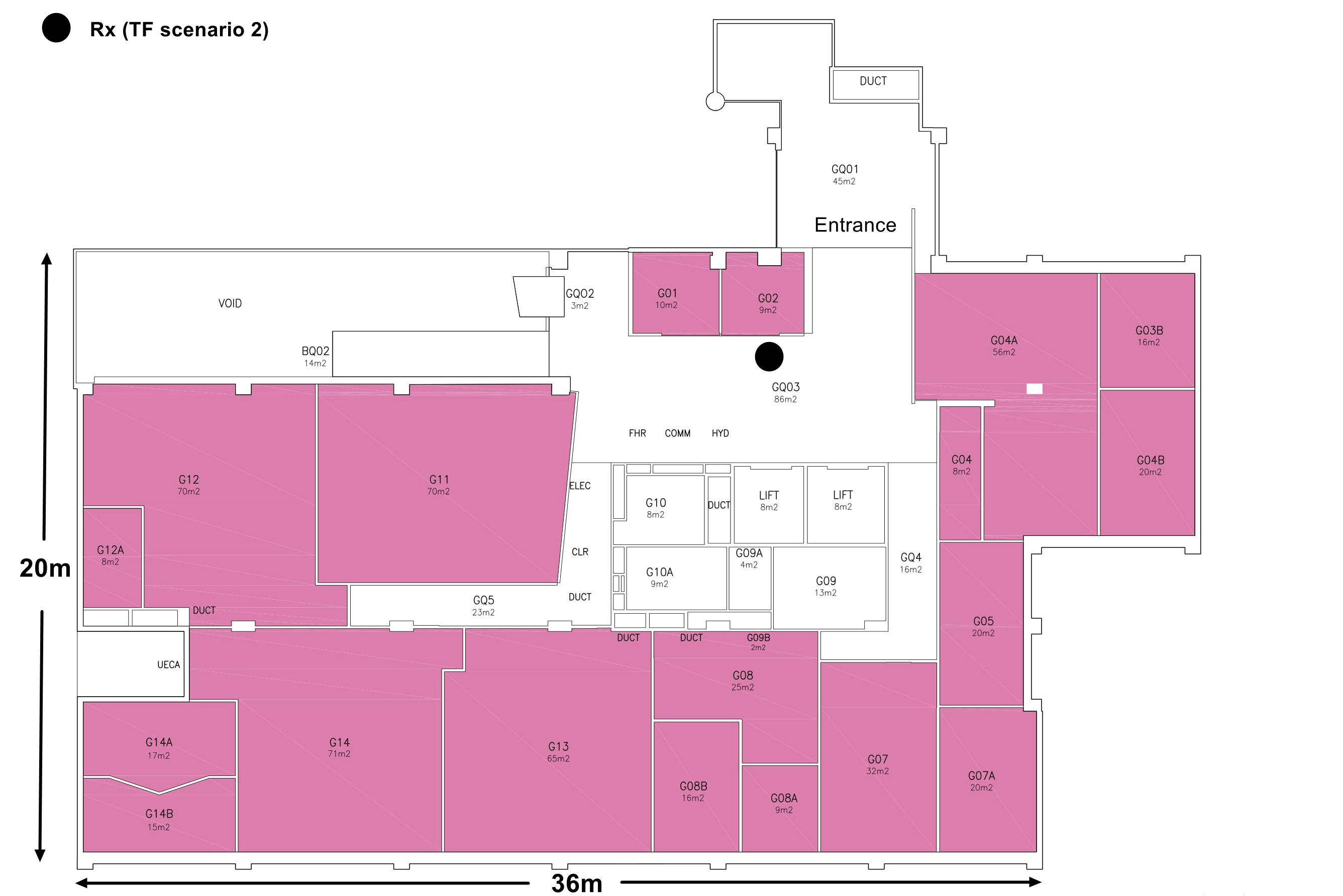}
		\label{fig:glevel}}
	\caption{Experimental setup.}
	\label{fig:setup}
	\vspace{-0.2in}
\end{figure*}

\section{Measurement Configuration}
\label{config}
\subsection{Experimental Environment}
As the products of WBS Technology can be deployed in different types of buildings,  we conduct study in four different types of buildings: an office building, a residential building, a car park and a warehouse. Tab.~\ref{tab:details} summarizes the details of these buildings and Fig.~\ref{fig:setup} provides some pictures of these buildings. The office building is a reinforced concrete building, which is located on a university campus. The building has 6 floors and 1 basement. The floor plan of level 4 and ground level is shown in Fig.~\ref{fig:level4} and Fig.~\ref{fig:glevel}, respectively. The residential building has 5 floors and 3 basements while the car park has 5 aboveground levels. The warehouse has only 1 level but many metal shelves and boxes as shown in Fig.~\ref{fig:warehouse}. 

\subsection{Measurement Apparatus}
\label{subsec:apparatus}
The hardware platform used in the study is Multitech mDot~\footnote{https://www.multitech.com/brands/multiconnect-mdot} which comprises a LoRa wireless chip (SX1272), an ARM processor and the LoRaWAN protocol stack. For the office building, we have deployed a LoRa testbed which consists of 10 LoRa motes.  The LoRa mote is connected to a Raspberry Pi as shown in Fig.~\ref{fig:hardware}. The Raspberry Pi will collect log data sent from LoRa mote through USB port, then transmit the data to a local server via Ethernet. The LoRa motes are evenly distributed in the building. Each mdot is equipped with an omni-directional vertically polarized antenna with gain of 3dBi.

For the rest of testing buildings, we use two mDots to collect data: they are configured as transmitter (Tx) and receiver (Rx), respectively. The RSSI reported by mDot is just an indication (represented by a number) of the power level being received by the antenna. Thus, a calibration of the LoRa mote using the spectrum analyser has been performed to determine the shift constant between the RSSI and the radio-frequency (RF) power. The method to determine the relation between the RSSI values and the real radio power is the same as~\cite{benaissa2016experimental}. A constant shift of 2 dB has been found between the RSSI reported by mDot and the RF power measured by spectrum analyser. For all the experiments in this paper, the parameters of LoRa module are set to the default values in Tab.~\ref{tab:differentparameter} unless otherwise stated.
\vspace{-0.1in}
\linespread{1.1}
\begin{table}[!ht]
\tiny
\centering
\caption{Details of experimental environment.}
\label{tab:details}
\begin{tabular}{|c|c|c|c|c|}
\hline
Building No & Type            & Size ($m^{3}$) & \begin{tabular}[c]{@{}c@{}}No. of\\ Floors\end{tabular} & \begin{tabular}[c]{@{}c@{}}No. of\\ Basement\end{tabular} \\ \hline
1 &Office Building      & $20\times36\times27$ & 6             & 1               \\ \hline
2 &Residential Building & $45\times55\times22$ & 5             & 3               \\ \hline
3 &Car park             & $65\times70\times18$ & 5             & 0               \\ \hline
4 &Warehouse             & $16\times60\times10$ & 1             & 0               \\ \hline
\end{tabular}
\vspace{-0.2in}
\end{table}
\linespread{1}

\section{Measurement Methodology}
\label{sec:method}
\subsection{Characterization of large-scale fading}
\label{sub:pathloss}
Large-scale fading is defined as the variability of received power with distance. The parameter characterizing large-scale fading is the path loss (PL). 
\subsubsection{Measurement Method}
\begin{figure}[!h]
	\centering
		\includegraphics[width=3in]{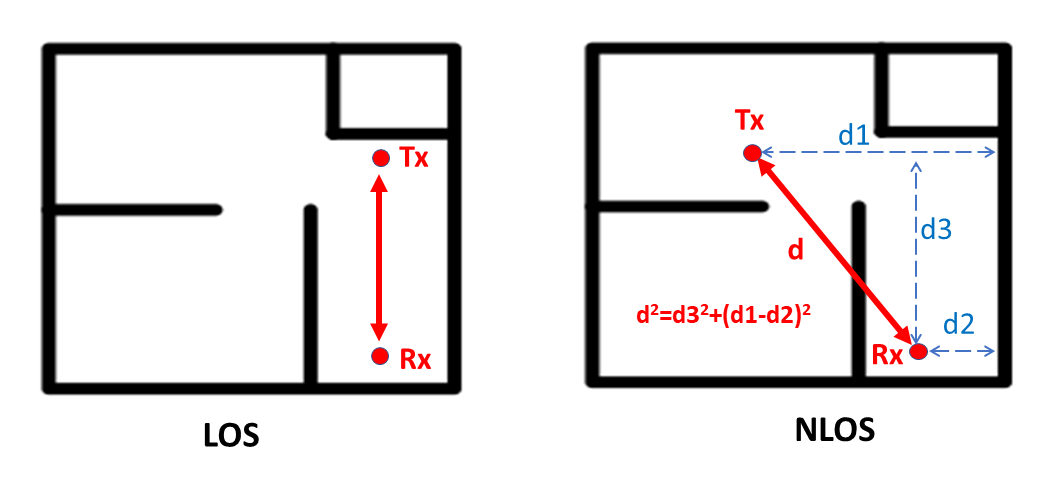}
	\caption{Distance measure.}
	\label{fig:distance}
	\vspace{-0.2in}
\end{figure}

To determine large-scale fading properties of LoRa in the indoor environment, two LoRa nodes are used in the data collection phase: Tx and Rx. The Tx is placed at a fixed position which is 1.5m above floor. The Rx is placed at different locations to collect RSSI at different distances. The distance between Tx and Rx is measured by a laser meter measurer. When there is LOS between Tx and Rx, the distance can be measured directly using a laser meter measurer. When there is NLOS between Tx and Rx, we calculate the distance of Tx and Rx to a fixed point, then the distance between Tx and Rx can be measured by pythagorean theorem as depicted in Fig.~\ref{fig:distance}.

For each building, the measurement is conducted in three different Large-scale fading (LSF) scenarios. The LSF scenarios are divided by whether or not LOS between the Tx and Rx exists, and by the number of floor between them. LSF scenario categories are specified as follows. Due to space limitation, we do not plot the floor plan of each building in this paper. Instead, we only plot the floor plan and experimental settings of the office building for illustrative purpose.
\begin{itemize}
\item \textit{LSF scenario 1: line-of-sight (LOS) path.} LOS between the Tx and Rx exists at every point along the path. For example, in the office building, the path is along a straight aisle on the fourth floor as shown in Fig.~\ref{fig:level4} (blue line). 
\item \textit{LSF scenario 2: obstructed line-of-sight path (OBS) on the same level.} Line of sight between the Tx and Rx is occasionally blocked between Tx and Rx. For example, in the office building, path is along an aisle other than the aisle where the Tx is located as shown in Fig.~\ref{fig:level4} (green line).
\item \textit{LSF scenario 3: none-line-of-sight (NLOS) path on different levels}. There is no LOS between the Tx and Rx. In the testing buildings which has more than 1 floor (i.e., office building, residential building, car park), the researcher walks along an aisle at different levels other than the level where the Tx is located. The experiment in the warehouse is conducted on the same floor as it has 1 level only. In this case, there are multiple shelves between the Tx and Rx to make sure there is no LOS between them.
\end{itemize}

\subsubsection{Path Loss Model}
\label{subsubsec:pathloss}
Large-scale fading characteristics of the indoor radio channel are determined by measurements of the path loss. The path loss model can be used in the link budget calculation.
A distance dependent path-loss model (also called one-slope model) has been demonstrated to perform well in indoor environments~\cite{rappaport1996wireless,aragon2008antennas}. According to this empirical model, the relationship between path loss $PL(d)$ in dB and distance $d$ in $m$ between the Tx and Rx can be expressed as follows:
\begin{equation}
PL(d)=PL(d_{0})+10nlog(\frac{d}{d_{0}})+X_{\sigma}
\label{eq:general}
\end{equation}
where $PL(d_{0})$ is the path loss at reference distance (1m in our measurement), $n$ is the path loss exponent, $d$ is the separation distance between the TX and RX, and $X_{\sigma}$ is a zero-mean Gaussian distributed variable (in dB) with standard deviation $\sigma$, also in dB. 

In this experiment, path loss samples are gathered up to a distance of 70m between the Tx and Rx. The parameters $PL(d_{0})$ and $n$ are obtained by fitting the path loss model to the measured path loss samples in a least-squares sense. Following~\cite{erceg1999empirically,tanghe2008industrial}, the intercept $PL(d_{0})$ can be determined in two ways:
\begin{itemize}
\item \textit{non-fixed intercept:} the intercept $PL(d_{0})$ is considered as a separate outcome of the least-squares fit.
\item \textit{fixed intercept:} the intercept $PL(d_{0})$ is chosen fixed and equal to the path loss at reference distance $d_{0}$ which is 1m in our measurement.
\end{itemize}
In this paper, a comparison between non-fixed and fixed intercept method is made to investigate which approach provides a better fit to measurement data.
\subsubsection{Shadowing effect}
\label{subsub:shadowing}
To characterise shadowing properties of the indoor environment, deviations $X_{\sigma}$ in dB between the path loss model with estimated parameters and the measured PL samples are calculated by $X_{i} = PL(d_{i}) - PL_{i}$,
where $PL_{i}$ is the $i$-th path-loss sample measured at distance $d_{i}$ and $PL(d_{i})$ is the path loss value predicted by the empirical model at distance $d_{i}$. It is shown that shadow fading samples $X_{i}$ closely follow a lognormal distribution with median equal to 0 dB~\cite{rappaport1996wireless,aragon2008antennas}.
For a particular path, shadow fading samples $X_{i}$ are calculated with respect to the path loss model fitted to the measurements collected along that path. Suppose we obtain $N$ shadow fading samples from a measurement track, the normalized autocorrelation function $Rxx (i)$ associated with that track is calculated as: $Rxx(i)=\frac{\sum{}^{N-i}_{p=1}X_{p}X_{p+i}}{\sum{}^{N}_{q=1}X^{2}_{q}}$.

For each track, we calculate the autocorrelation function to analyse the shadowing characteristics. Particularly, we also calculate the decorrelation distance which is defined as the distance to which the normalized autocorrelation drops below 0.1. This definition is commonly used, as autocorrelation can often be modeled as exponentially decaying with distance~\cite{jalden2007correlation}.  
\subsection{Characterization of Temporal Fading}
\label{sub:temporal}
Temporal fading (TF) is defined as the variability of received signal strength over time at a fixed location in the propagation environment. 
Previous studies show fading statistics follows a Rician distribution when a dominant multipath component (e.g., LOS component) exists~\cite{benaissa2016experimental,parsons1992mobile}. Therefore, we model the measured temporal fading samples and compare with the theoretical Rician distribution.
\subsubsection{Measurement Method}
\label{subsub:method}
Different from large-scale fading, the temporal fading is not determined by stationary physical characteristics such as LOS or NLOS. Instead, it is caused by movement of persons in the multipath environment. To determine temporal fading properties of LoRa technology in the multi-floor building, two different TF scenarios are considered:
\begin{itemize}
\item \textit{TF scenario 1:} Both the Tx and Rx are put at fixed location on the same floor. As an example, the positions of Tx and Rx are shown in Fig.~\ref{fig:level4}.
\item \textit{TF scenario 2:} The Tx and Rx are located at different levels. For example, in the office building the Tx remains at level 4 while the Rx is put at a fixed position on ground level as shown in Fig.~\ref{fig:glevel}. In the warehouse, the Tx and Rx are separated by a number of shelves.
\end{itemize}

In each scenario, the received signal strength are recorded in a time span of 1hr, at a rate of approximately 30 samples per second. After data collection, the median received power $P_{median}$ in dBm is removed from the received power samples $P_{i}$ as $Y_{i}=P_{i}-P_{median}$.
The samples $Y_{i}$ will be compared with the Rician distribution to analyse the temporal fading characteristics. The Rician distribution is often described in terms of a parameter $K$ (Rician factor), which is defined as the ratio between the power received via the dominant path and the power contribution of the obstructed paths~\cite{abdi2001estimation}. The parameter $K$ is given by $K=A^2/2b^2$ or in terms of dB $K=10log(\frac{A^2}{2b^{2}})$,
where $A^2$ is the energy of the dominant path and $2b^2$ is the energy of the diffuse part of the received signal~\cite{abdi2001estimation}. From the definition of the Rician K-factor, low K-factor indicates large motion (i.e., large b) within the wireless propagation environment that disturbs the received power profile over time, while large K-factor reveals a low movement in the environment. To estimate the K-factor, two distinct methods are used in this study:
\begin{itemize}
\item \textit{moment-based estimator:} to estimate the K factor, the method of moments proposed in~\cite{abdi2001estimation} is used. This method provides a simple parameter estimator based on the variance and the mean of the received signal strength. 
\item \textit{curve fitting:} To estimate the K-factor, the empirical cumulative distribution function (CDF) is constructed and compared to a Rician distribution with zero median in dB using a least-squares curve fitting technique. 
\end{itemize}
In this paper, the moment-based and the curve fitting method will be compared to determine how well temporal fading fits Rician distribution. The results of temporal fading are presented in Sec.~\ref{sec:temporalfadingresults}.
\subsection{Coverage Evaluation}
\label{sub:performance}
In this set of experiment, we aim to evaluate the coverage of LoRa technology in multi-floor buildings using different parameter settings. Specifically, we will focus on packet reception rate (PRR) as it is an important metric for wireless sensor network. The parameters investigated in this study include data rate, bandwidth (BW), center frequency, and spread factor (SF). During the experiment, we put Tx at top floor, and change the position of Rx from top floor to the lowest floor to investigate the changes in PRR. For each floor, we choose 10 evenly distributed points to collect data. For each point, the Rx keeps receiving packets sent from Tx for 30 minutes. The PRR of each floor is obtained by averaging the results of 10 points. As warehouse has only 1 level, we do not conduct experiment in warehouse. Tab.~\ref{tab:differentparameter} lists all the parameter values tested in this study. The parameter values marked in bold are called default values. 
\begin{table}[!ht]
\tiny
\centering
\caption{Parameter settings in coverage evaluation.}
\label{tab:differentparameter}
\begin{tabular}{|cccc|}
\hline
\begin{tabular}[c]{@{}c@{}}BW\\ (Khz)\end{tabular} & \begin{tabular}[c]{@{}c@{}}Center frequency\\ (Mhz)\end{tabular} & SF & \begin{tabular}[c]{@{}c@{}}Tx power\\ (dBm)\end{tabular} \\ \hline
 \textit{\textbf{500}}                                                & \textit{\textbf{915}}                                                              & \textit{\textbf{7}}  & \textit{\textbf{20}}                                                       \\ \hline
 250                                                & 919                                                              & 8  &                                                        \\ \hline
       125                                              & 923                                     & 9  &                                                        \\ \hline
                & 928                   & 10 &                                                        \\ \hline
\end{tabular}
\end{table}

\section{Large-scale Fading Results}
\label{sec:largescaleresults}
\linespread{1.1}
\begin{table}[!th]
\small
\centering
\caption{Non-fixed method VS fixed method.}
\label{tab:oneslope}
\resizebox{3.5in}{!}{
\begin{tabular}{|c|c|ccc|ccc|}
\hline
\multirow{2}{*}{Building}                                                       & \multirow{2}{*}{LSF Scenario} & \multicolumn{3}{c|}{Non-fixed Intercept} & \multicolumn{3}{c|}{Fixed Intercept} \\
                                                                                &                               & PL($d_{0}$) {[}dB{]}  & n {[}-{]}  & $\sigma${[}dB{]} & PL($d_{0}$) {[}dB{]} & n {[}-{]} & $\sigma${[}dB{]}      \\ \hline
\multirow{4}{*}{1}                                                         & 1-LOS                         & 38               & 2.17       & 4.99     & 37              & 2.35      & 5.04              \\ \cline{2-2}
                                                                                & 2-OBS                         & 39               & 2.43       & 5.17     & 37              & 2.54      & 5.18            \\ \cline{2-2}
                                                                                & 3-NLOS                       & 45               & 6.03        & 5.88     & 37              & 8.2      & 5.95              \\ \cline{2-8} 
                                                                                & Average                       & \textbf{40.7}           & \textbf{3.54}          & \textbf{5.34}             & \textbf{37}          & \textbf{4.36}        & \textbf{5.38}            \\ \hline
\multirow{4}{*}{2} & 1-LOS                         & 38          & 2.31          & 4.87             & 37          & 2.35        & 4.97            \\ \cline{2-2}
                                                                                & 2-OBS                         & 39          & 2.05          & 4.62             &37          & 4.71        & 4.84            \\ \cline{2-2}
                                                                                & 3-NLOS                        & 42          & 5.64          & 5.01             & 37          & 5.8        & 5.31            \\ \cline{2-8} 
                                                                                & Average                       & \textbf{39}           & \textbf{3.33}          & \textbf{4.83}            & \textbf{37}          & \textbf{4.29}        & \textbf{5.04}            \\ \hline
\multirow{4}{*}{3}                                                       & 1-LOS                         & 38          & 1.52          & 4.7             & 36          & 1.74        & 4.8            \\ \cline{2-2}
                                                                                & 2-OBS                         & 40          & 1.87          & 4.57             & 36          & 1.94        & 4.79            \\ \cline{2-2}
                                                                                & 3-NLOS                        & 46          & 8.4          & 5.32             & 36          & 8.7        & 5.64            \\ \cline{2-8} 
                                                                                & Average                       & \textbf{41.3}           & \textbf{3.93}          & \textbf{4.86}             & \textbf{36}          & \textbf{4.13}        & \textbf{5.07}            \\ \hline
\multirow{4}{*}{4}                                                      & 1-LOS                         & 42          & 1.7          & 5.23             & 36          & 2.43        & 5.85            \\ \cline{2-2}
                                                                                & 2-OBS                         & 43          & 1.74          & 5.11             & 36          & 2.27        & 5.47            \\ \cline{2-2}
                                                                                & 3-NLOS                        & 44          & 3.8          & 5.49             & 36          & 4.62        & 5.92            \\ \cline{2-8} 
                                                                                & Average                       & \textbf{42.7}           & \textbf{2.41}          & \textbf{5.27}             & \textbf{36}          & \textbf{3.11}        & \textbf{5.74}            \\ \hline
\end{tabular}
}
\end{table}
\linespread{1}
\subsection{Path loss per LSF scenario}
\label{subsec:pathloss}
The parameters of the one-slope model in Sec.~\ref{subsubsec:pathloss} are determined separately for each LSF scenario. Tab.~\ref{tab:oneslope} shows the path loss $PL(d_{0})$ at reference distance $d_0 = 1~m$ and the path-loss exponent $n$, as well as the standard deviation $\sigma$ of the samples. Distinction has been made between one-slope models with non-fixed and fixed intercept $PL(d_0)$. In the following, results of the path-loss measurements are discussed.
\begin{figure*}[!ht]
	\centering
	\subfigure[LSF scenario 1]{
		\includegraphics[width=1.8in]{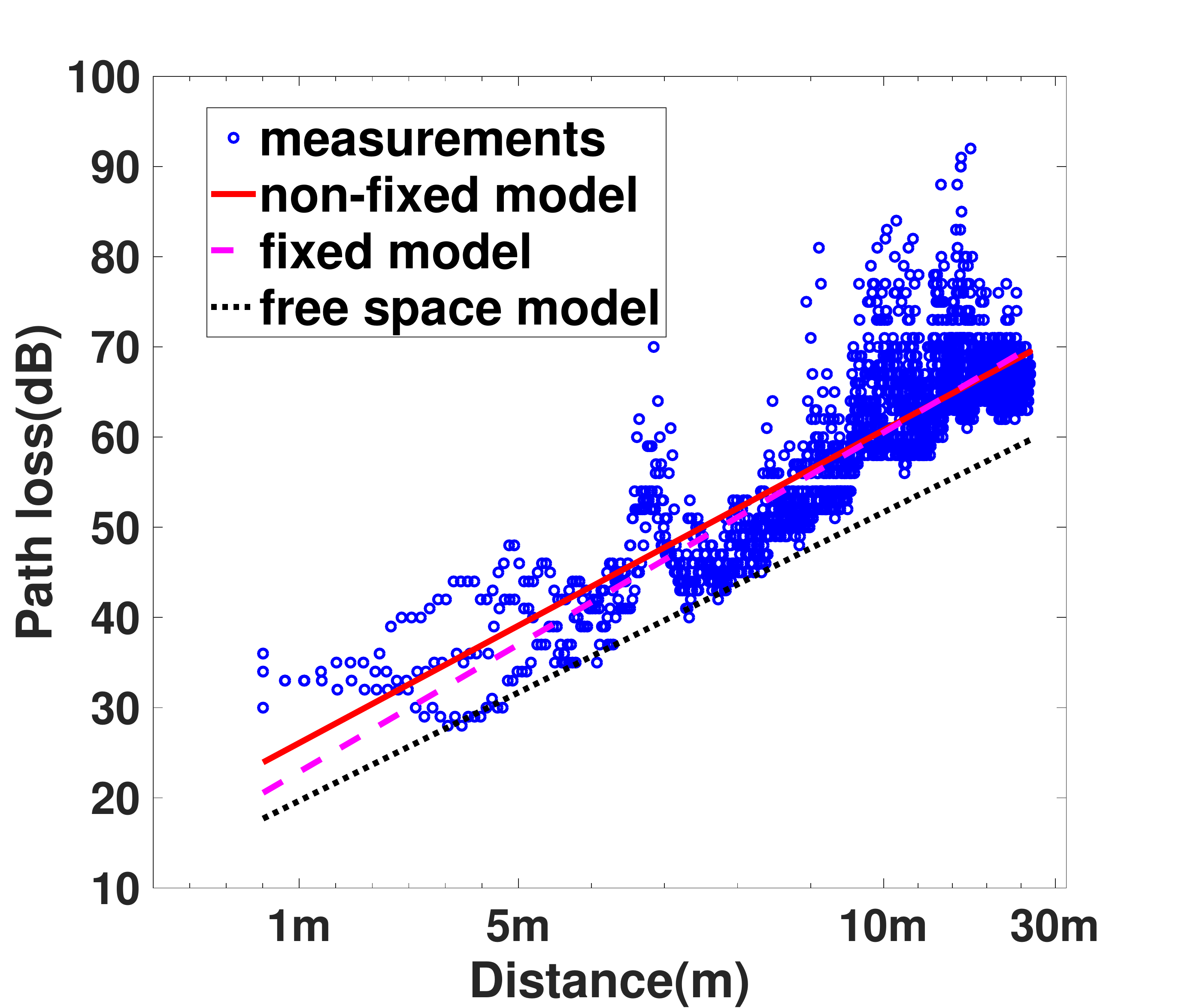}
		\label{fig:samefloorLOS}}
		\vspace{-0.01in}
	\subfigure[LSF scenario 3 (ground level)]{
		\includegraphics[width=1.8in]{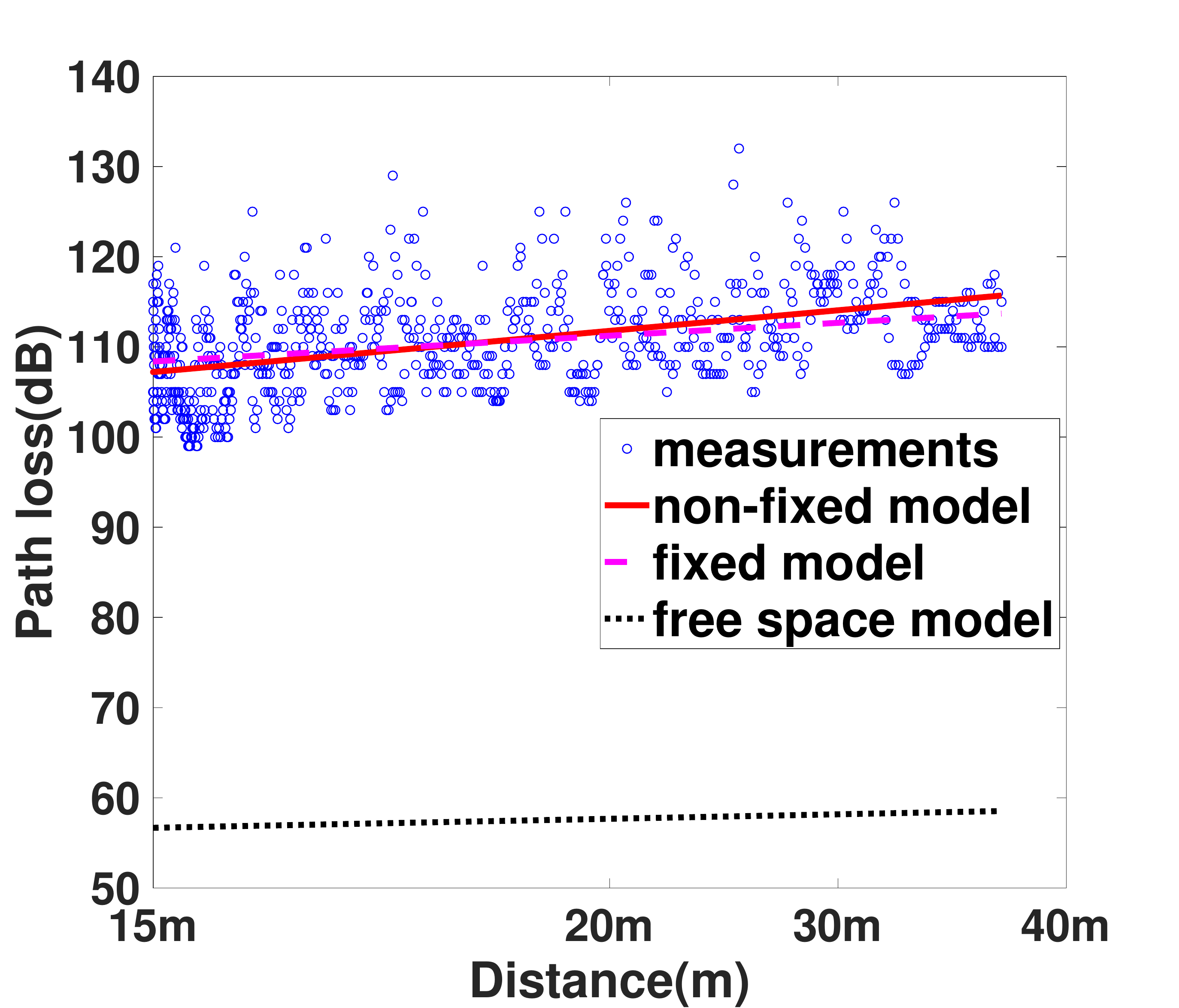}
		\label{fig:4floors}}
		\vspace{-0.01in}
		\subfigure[Decorrelation distance]{
		\includegraphics[width=1.8in]{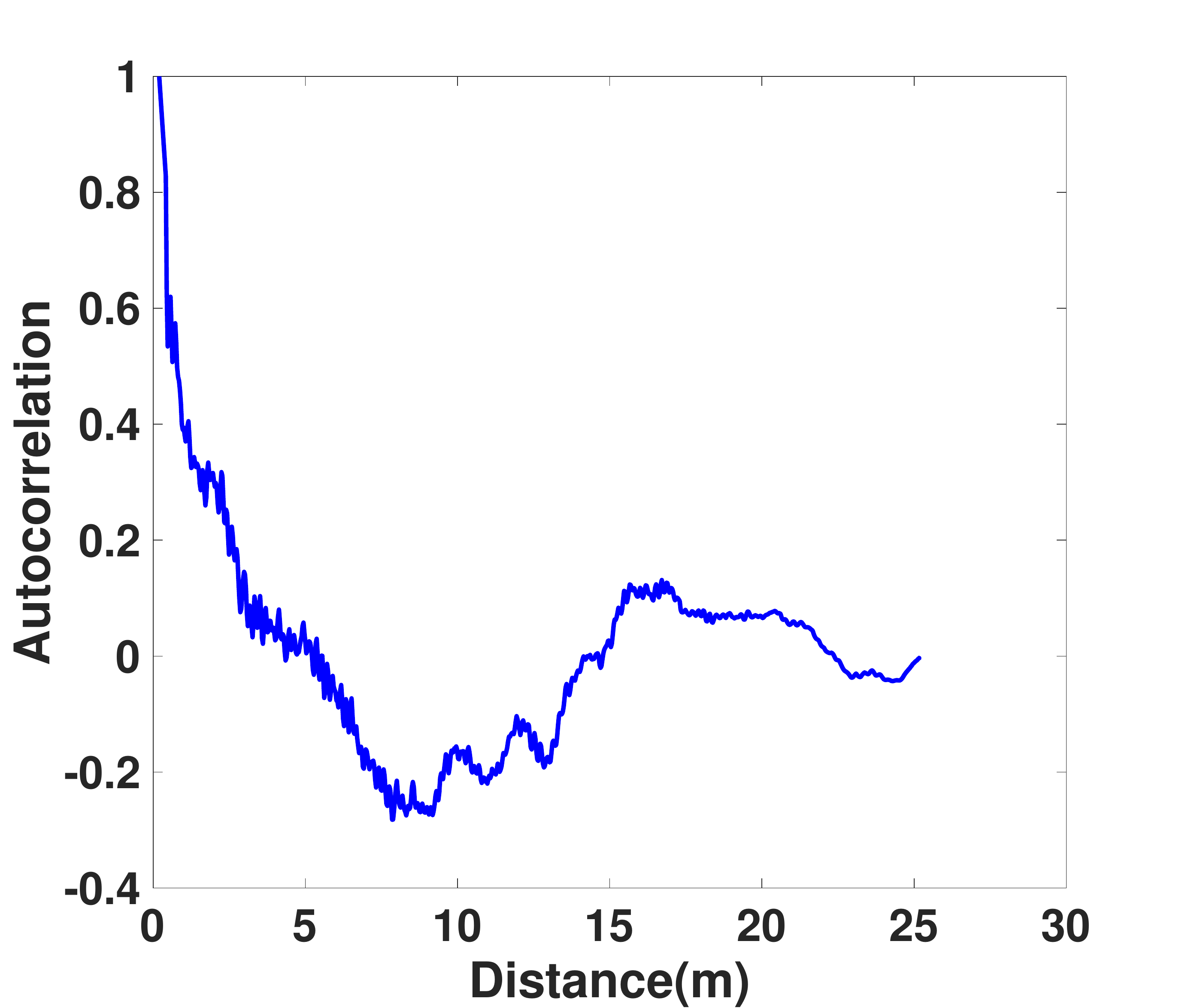}
		\label{fig:autocorrelation1}}
		\vspace{-0.01in}
		\caption{Evaluation results of large scale fading.}
		\label{fig:LSFevaluationresults}
		\vspace{-0.2in}
\end{figure*}
\paragraph{\textbf{Comparison between non-fixed and fixed intercept models.}}
The results of different buildings and different methods are summarized in Tab.~\ref{tab:oneslope}. From Tab.~\ref{tab:oneslope}, we find some regular patterns. First, for all the buildings, the intercept $PL(d_0)$ of non-fixed intercept model is consistently larger than that of fixed intercept model. Second, smaller path loss exponents $n$ are obtained with the non-fixed intercept model in comparison with the fixed intercept model. Third, standard deviations $\sigma$ for non-fixed intercept and fixed intercept match well, although they are consistently somewhat larger for the fixed intercept method.

Based on the smaller stand deviation between the model and real measurements, we can draw the conclusion that the use of a path-loss model with non-fixed intercept is superior to using a fixed-intercept model. This is due to the following two reasons: firstly, it is noteworthy that the difference in $PL(d_0)$ between the non-fixed intercept and the fixed intercept models in Tab.~\ref{tab:oneslope} can occasionally be as high as 8 dB. Therefore, free-space propagation at reference distance $d_0$ (assumed for fixed intercept models) can hardly be assumed for our measurements. This is probably caused by the multi-path effect of the indoor environment. Free-space propagation can usually only be assumed in spacious areas free of immediate obstructions, in contrast to our measurement configuration where the Tx and Rx are obstructed by soft partitions and floors.

Secondly, it is generally accepted in literature that shadow fading samples $X_i$ are log-normally distributed. To ascertain log-normality, a Kolmogorov-Smirnov (K-S) goodness-of-fit test has been performed on the shadow fading samples of the one-slope models provided in Tab.~\ref{tab:oneslope}. For the K-S test, the empirical CDF of shadow fading samples is compared to a log-normal CDF with zero median in dB and standard deviation $\sigma$ of the corresponding model. For the three LSF topographies in Tab.~\ref{tab:oneslope}, the non-fixed intercept one-slope models passed the K-S test at $\alpha = 0.05$ level of significance, whereas only three out of twelve fixed intercept models passed the same test. This indicates that median path loss is most accurately specified by a path-loss model with non-fixed intercept. Moreover, standard deviations $\sigma$ in Tab.~\ref{tab:oneslope} are somewhat smaller for non-fixed intercept in comparison with fixed intercept, showing that a non-fixed intercept one-slope model provides a better fit to experimental path-loss data.

As an example, Fig.~\ref{fig:samefloorLOS} and~\ref{fig:4floors} shows the measured path-loss samples and the predicted model of LSF scenario 1 and ground level of LSF scenario 3, respectively. Also shown in the figure is the free-space path loss at 915~Mhz. As expected, the path loss predicted by both the non-fixed intercept and fixed intercept is higher than the free space path loss.

In Tab.~\ref{tab:oneslope}, we can also observe building-to-building path-loss variations. However, building-to-building variation is not large. For example, in terms of LSF scenario 1, the path loss exponents $n$ vary only from 1.54 to 2.43 among different types of buildings. This is mainly because construction details of all measured buildings are similar (e.g., concrete floors and walls).
\begin{figure*}[htb] 
    \centering 
    \subfigure[Same floor]{
		\includegraphics[width=3in]{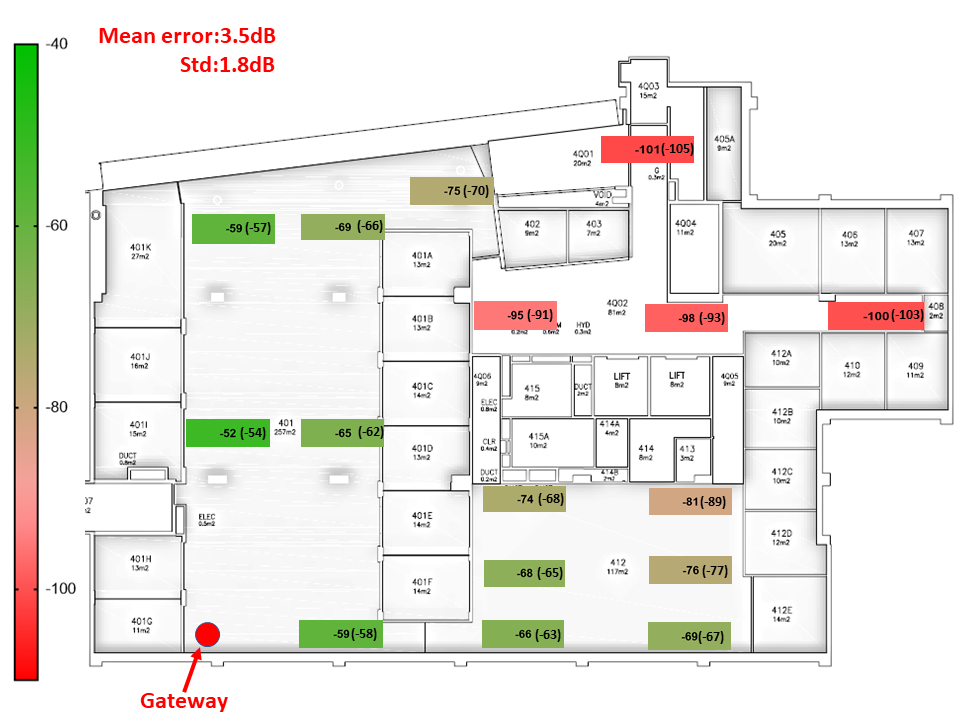}
		\label{fig:tool_horizontal}}
	\subfigure[Different floor]{
		\includegraphics[width=3in]{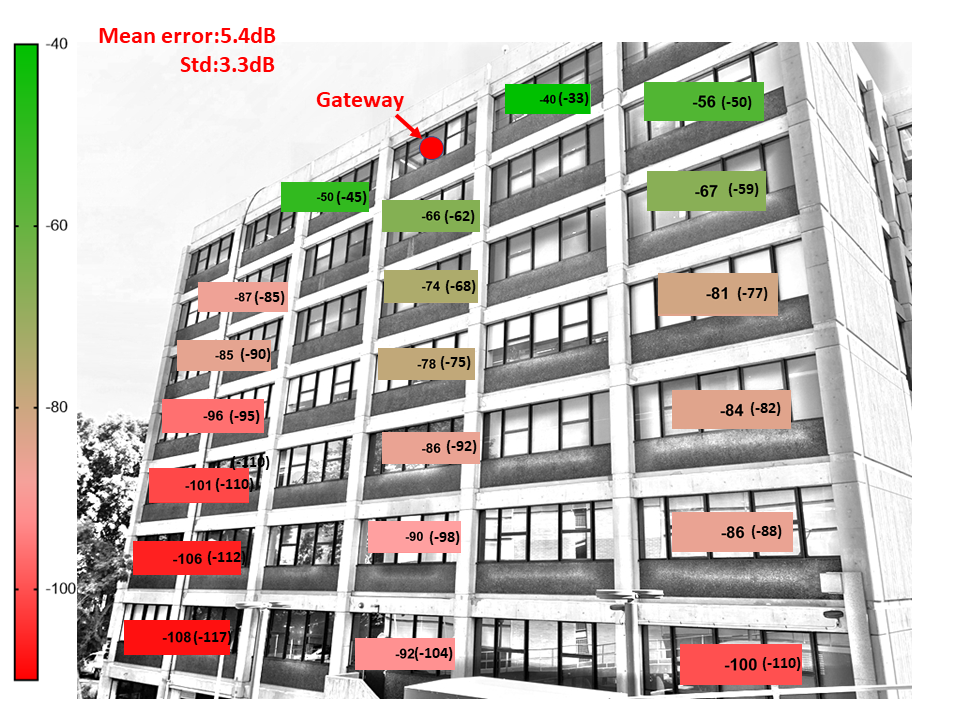}
		\label{fig:tool_vertical}}
	\caption{Accuracy of path loss model (the values in the parenthesis are predicted value).}
	\label{fig:toolchain}
	\vspace{-0.2in}
\end{figure*}

\paragraph{\textbf{Soft Partition and Concrete Wall Attenuation Factor Model}}
In the previous section, the path loss in multi-floored environment is predicted by a model that includes distance only. However, from Tab.~\ref{tab:oneslope} it can be seen that the path loss exponent $n$ changes greatly in different LSF scenarios and the stand deviation $\sigma$ can be as high as 5.74 dB. These parameters may be used in the model for a first-order prediction of mean signal strength when only distance but no other information such as the number of floors is known, but is clearly unsatisfactory for site layout or capacity prediction. 

There are often obstructions between the Tx and Rx such as soft partitions, walls and floors. In order to build a more accurate propagation model, we need to consider the path loss effects of these obstructions. In the literature~\cite{seidel1992914}, this is achieved by including the attenuation factor of floor, soft partition and wall in the prediction model. For simplicity, we assume that any kind of concrete support column that wholly or partially blocks the direct path between the Tx and Rx is labeled as a concrete wall. Let $p$ and $q$ be the number of soft partition and concrete wall between the Tx and Rx respectively. Then the path loss predicted by the attenuation factor model (AF model) is given by:
\begin{multline}
\begin{split}
PL(d)=&PL(d_{0})+10nlog(\frac{d}{d_{0}}) +FAF\left[dB\right]\\
&+p\ast AF_{partition}\left[dB\right]+q\ast AF_{wall}\left[dB\right]
\label{eq:pathlossmodel2}
\end{split}
\end{multline}
where FAF is the floor attenuation factor, $AF_{partition}$ and $AF_{wall}$ are the attenuation factor of one soft partition and concrete wall. To obtain a more precise path loss model, we conduct a drive test to calculate the attenuation factors of different types of obstructions in the test building. Specifically, we first measure the path loss when the Tx and Rx is separated by 2m, then we calculate the path loss when the Tx and Rx is separated by the same distance but with an obstruction between them (e.g, wall, soft partition). The difference of path loss between these two tests is regarded as the attenuation factor of this type of obstruction. For each type of obstruction, we repeat the test for multiple times at different locations of the building, and the final result is obtained by calculating the mean of these tests. It is worth mentioning that the attenuation factor are only calculated in the office building.

The attenuation factors of different types of obstruction are summarized in Tab.~\ref{tab:af}. For comparison purpose, we also list the attenuation factors reported in the literature. We can see that the FAF varies greatly in different buildings with varied frequencies. This result indicates that a site-specific path loss model is required because of the different construction materials and layout as well as many other factors. However, the attenuation factors of concrete wall, wooden door and soft partition in our measurements are similar to those reported in previous studies. The attenuation factor of glass measured in our study (2.04~dB) is different from previous results (4.5~dB). It is probably because the thickness of glass in~\cite{rath2017realistic} is different from that in our experimental building which is about 2cm. It is also interesting to note that the average FAF is not a linear function of the number of floors between the transmitter and receiver as also found in~\cite{keenan1990radio,motley1988personal}. It is possible that different floors cause different amounts of path loss, and there may be other factors such as multi-path reflections from surrounding buildings that affect the path loss. With the knowledge of attenuation factors, we re-calculate the path loss model using curve fitting with non-fixed intercept method. In order to demonstrate the benefits of AF model, in LSF scenario 3 we put TX on Level 4 and change the position of RX from Level 4 to Ground floor level by level.  From the results in Tab.~\ref{tab:comparison}, we find that the stand deviation reduces from 5.84 to 4.21 which indicates that the attenuation factor model is more accurate.

\linespread{1.1}
\begin{table}[!ht]
\centering
\small
\caption{Attenuation factors}
\label{tab:af}
\resizebox{3.3in}{!}{
\begin{tabular}{|ccccc|}
\hline
\multirow{2}{*}{Obstruction} & \multirow{2}{*}{AF (dB)} & \multicolumn{3}{c|}{Previous Studies} \\ \cline{3-5} 
                             &                                          & ~\cite{seidel1992914}(914Mhz)        & ~\cite{rath2017realistic}(2.4Ghz)         &~\cite{rath2017realistic}(2.4Ghz)        \\ \hline
1 floor                      & 21.7                                     & 12.9               & 16                 & 21                \\ 
2 floors                     & 25.9                                     & 18.7               & 27                 & 33                \\ 
3 floors                     & 27.2                                     & 24.4               & 31                 & 40                \\ 
4 floors                     & 50.8                                     & 27                 & -                  & -                 \\ 
Concrete wall                & 2.2                                      & -                  & -                  & 2.73              \\ 
Glass                        & 2.04                                     & -                  & -                  & 4.5               \\ 
Wooden door                  & 2.11                                     & -                  & -                  & 2.67              \\ 
Soft partition board         & 2.5                                      & -                  & -                  & 2.3 (2.5Ghz~\cite{anderson2004building})              \\ \hline
\end{tabular}
}
\end{table}
\begin{figure*}[!ht]
	\centering
	\subfigure[TF scenario 1]{
		\includegraphics[width=1.6in]{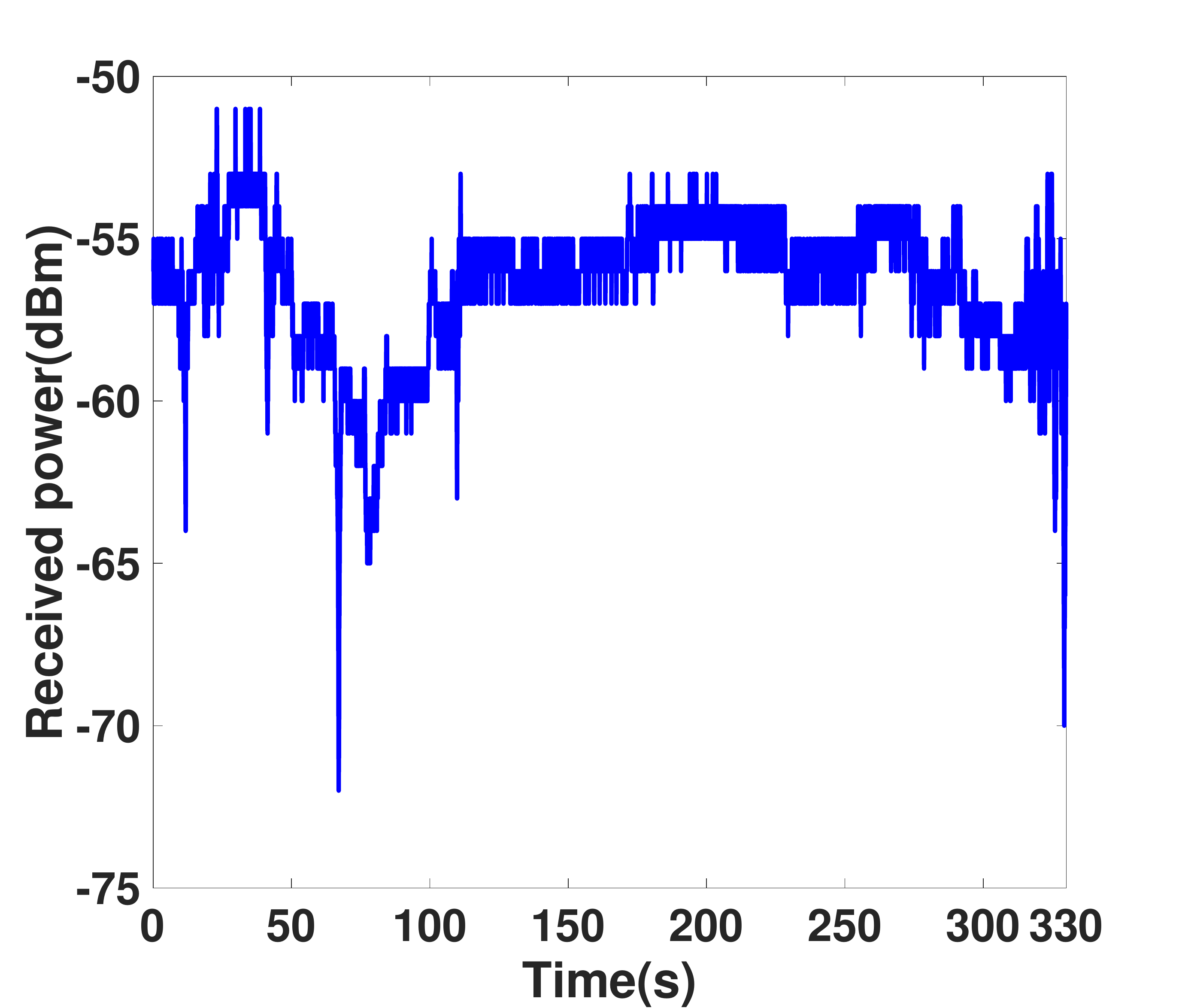}
		\label{fig:temporalfading1}}
	\subfigure[TF scenario 2]{
		\includegraphics[width=1.6in]{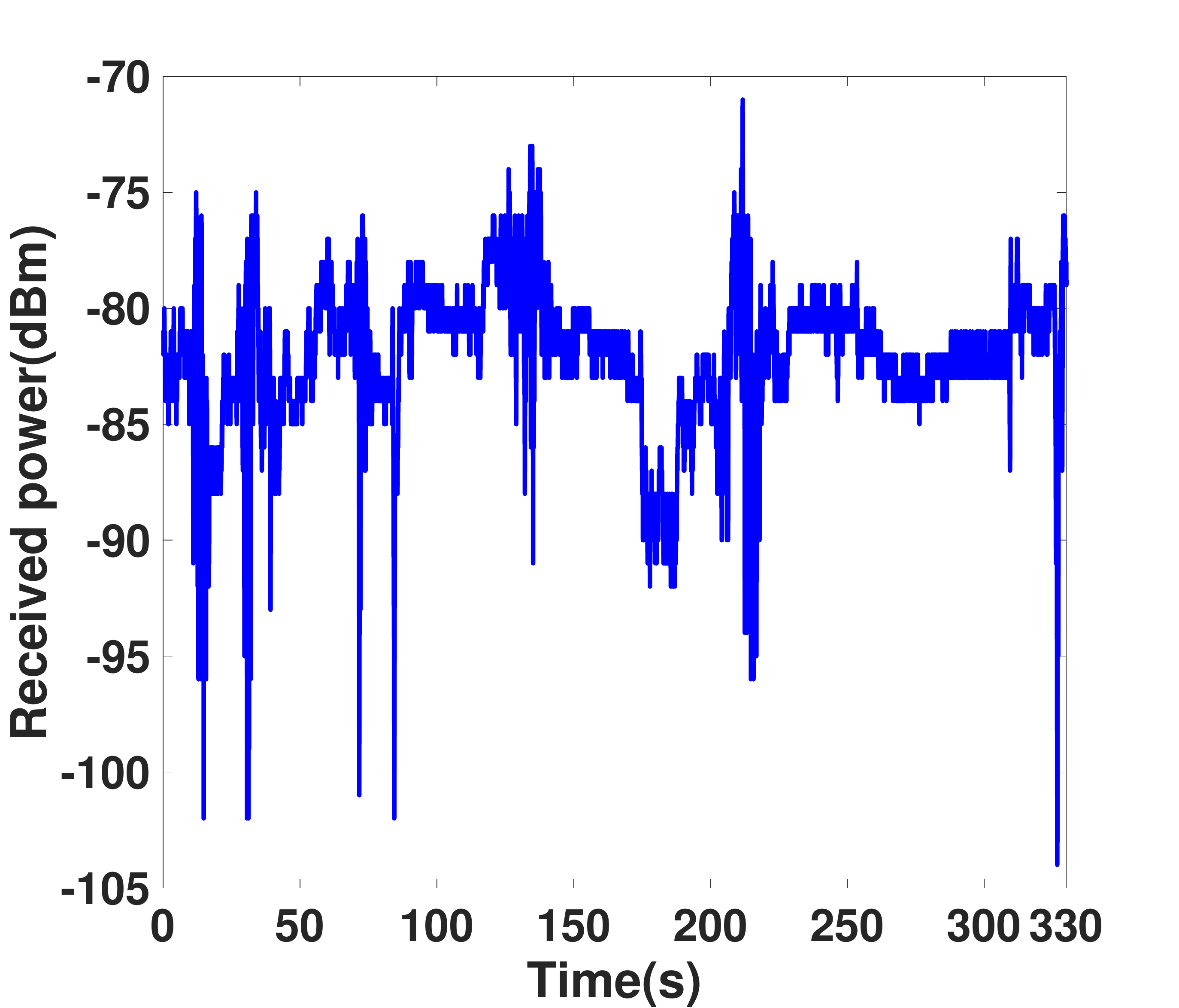}
		\label{fig:temporalfading2}}
       \subfigure[TF scenario 1-no people]{
		\includegraphics[width=1.6in]{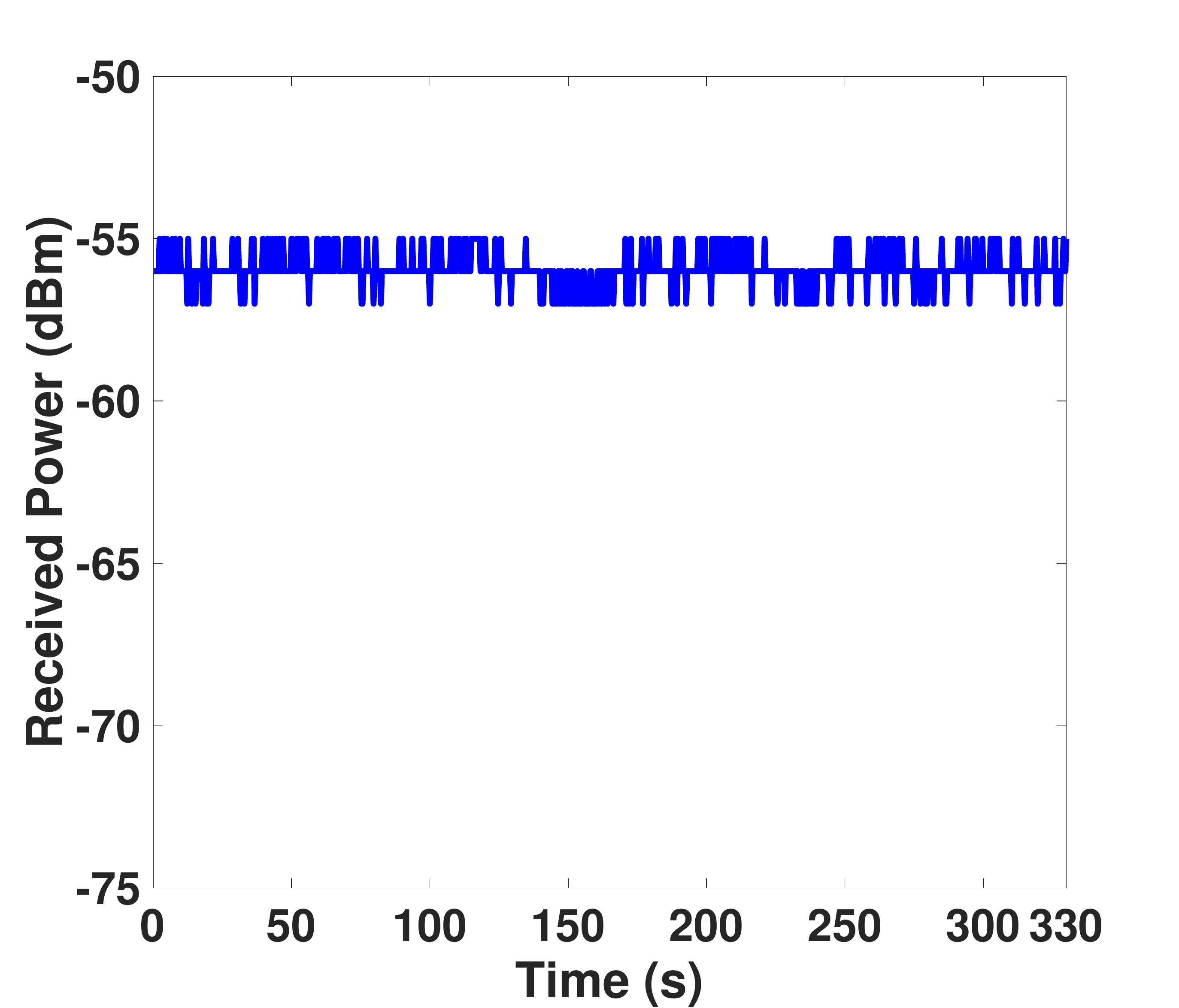}
		\label{fig:temporalfading3}}
	\subfigure[TF scenario 2-no people]{
		\includegraphics[width=1.6in]{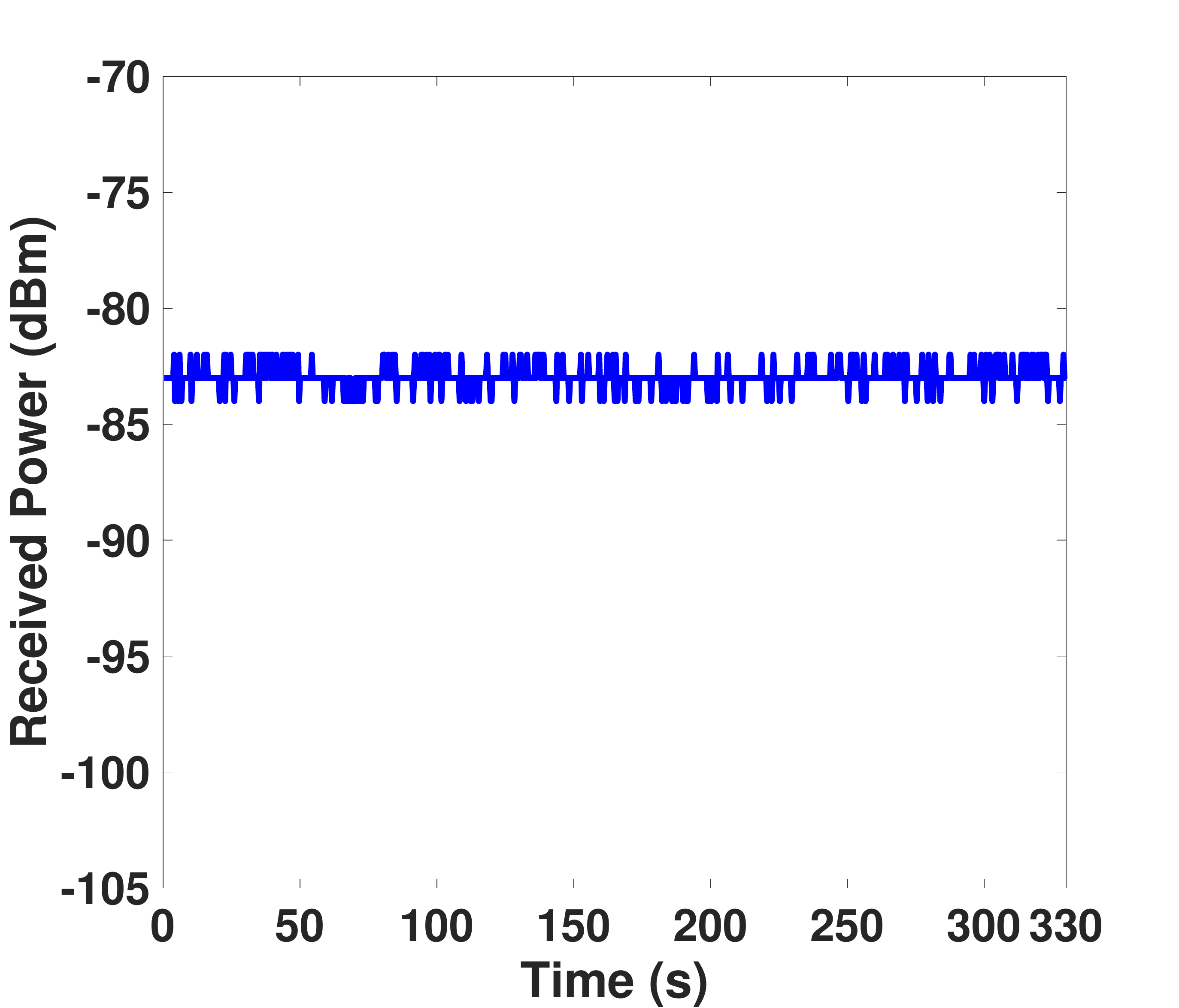}
		\label{fig:temporalfading4}}
	\caption{Temporal-fading measurement.}
	\label{fig:temporalfading}
	\vspace{-0.2in}
\end{figure*}
\begin{figure*}[!ht]
	\centering
	\subfigure[CDF of scenario 1]{
		\includegraphics[width=1.6in]{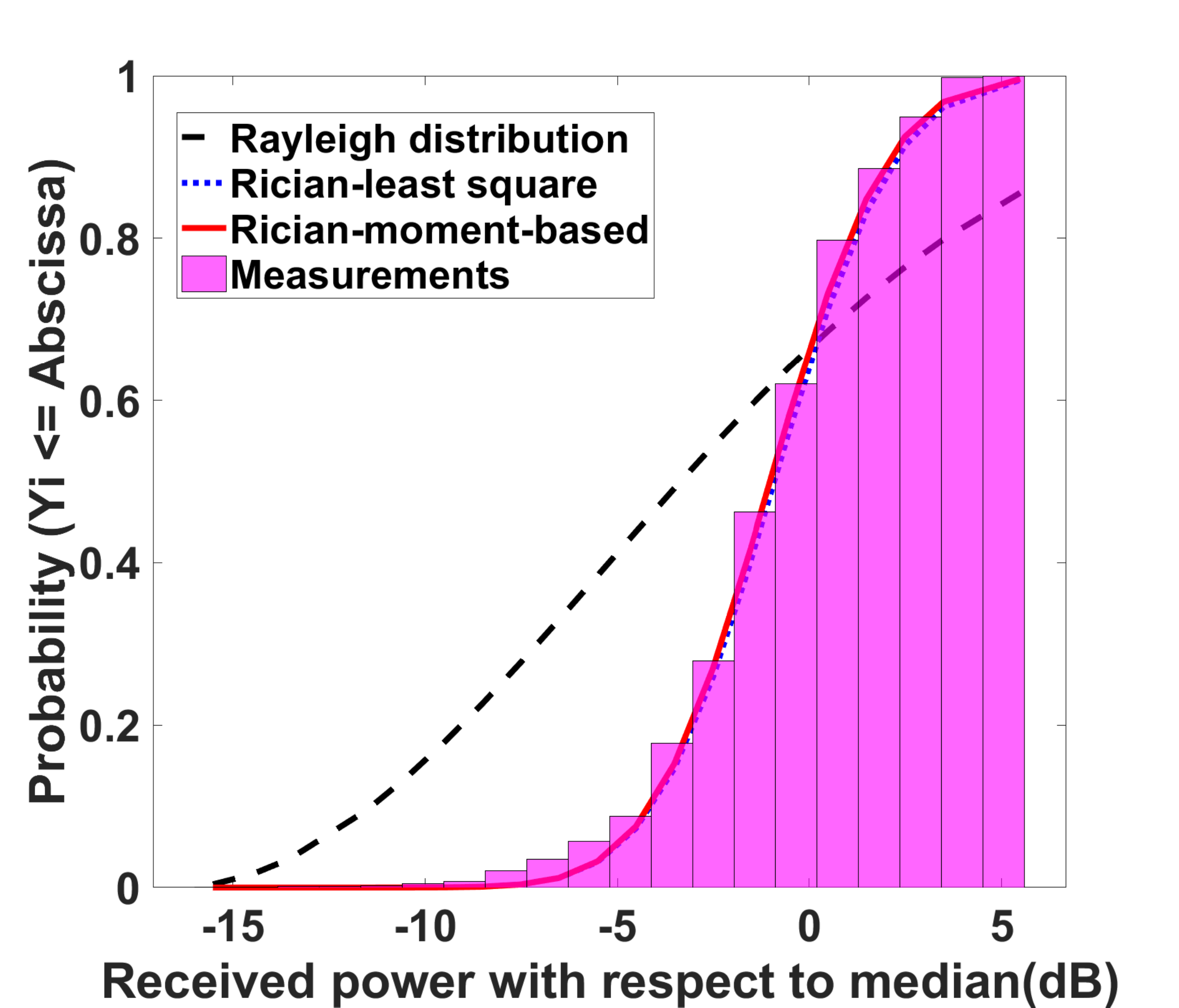}
		\label{fig:CDF1}}
	\subfigure[PDF of scenario 1]{
		\includegraphics[width=1.6in]{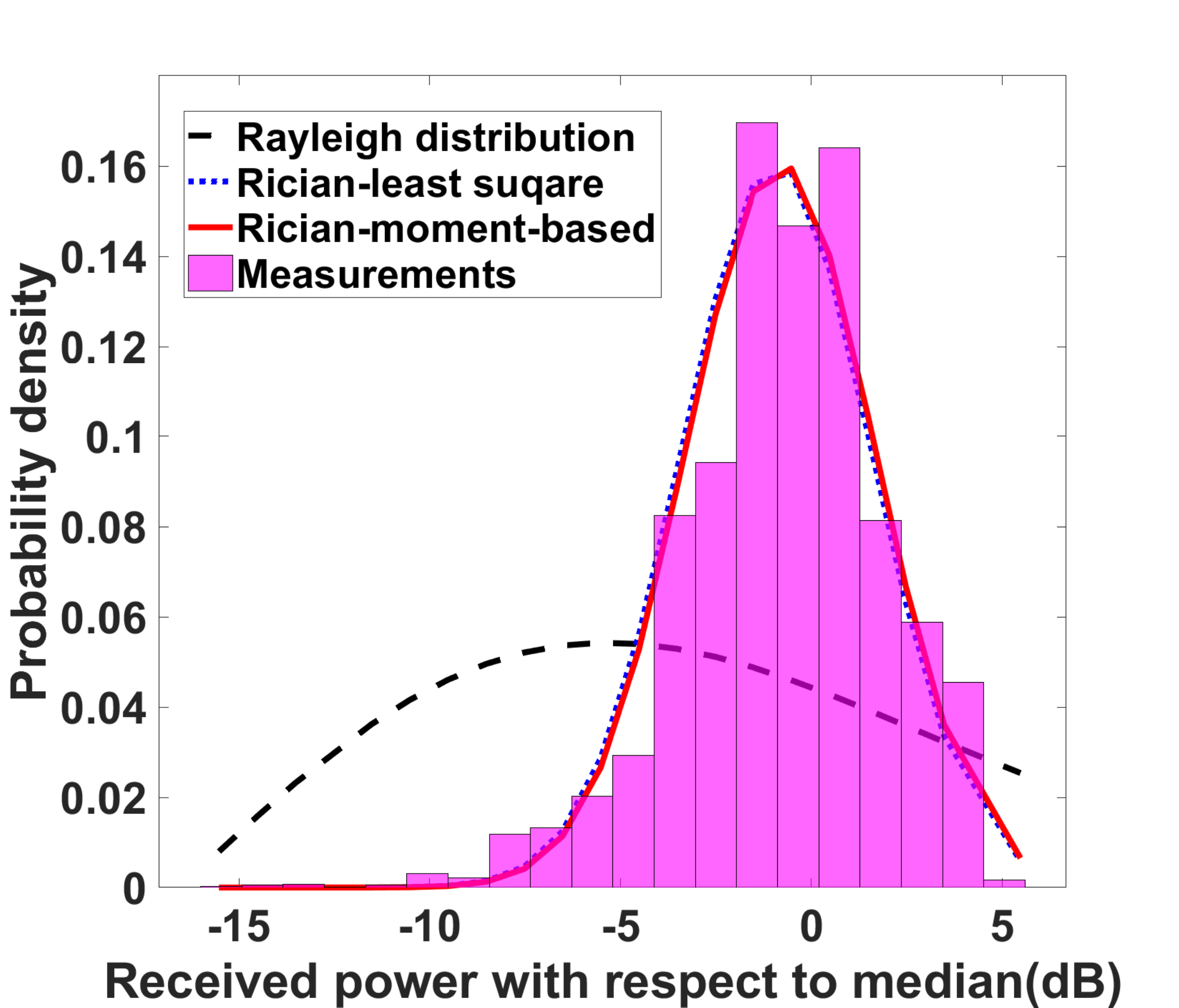}
		\label{fig:PDF1}}
		\subfigure[CDF of scenario 2]{
		\includegraphics[width=1.6in]{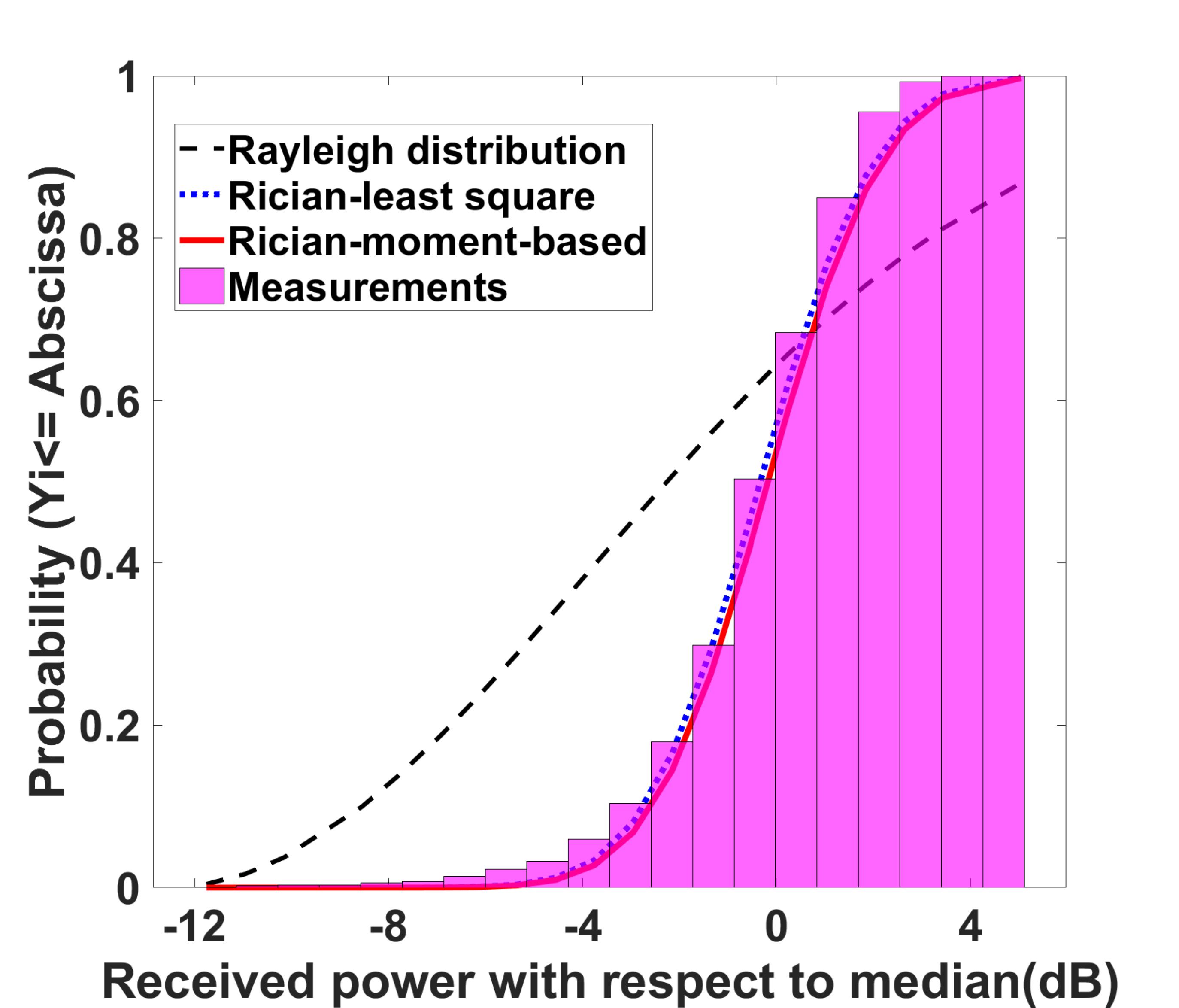}
		\label{fig:CDF2}}
	\subfigure[PDF of scenario 2]{
		\includegraphics[width=1.6in]{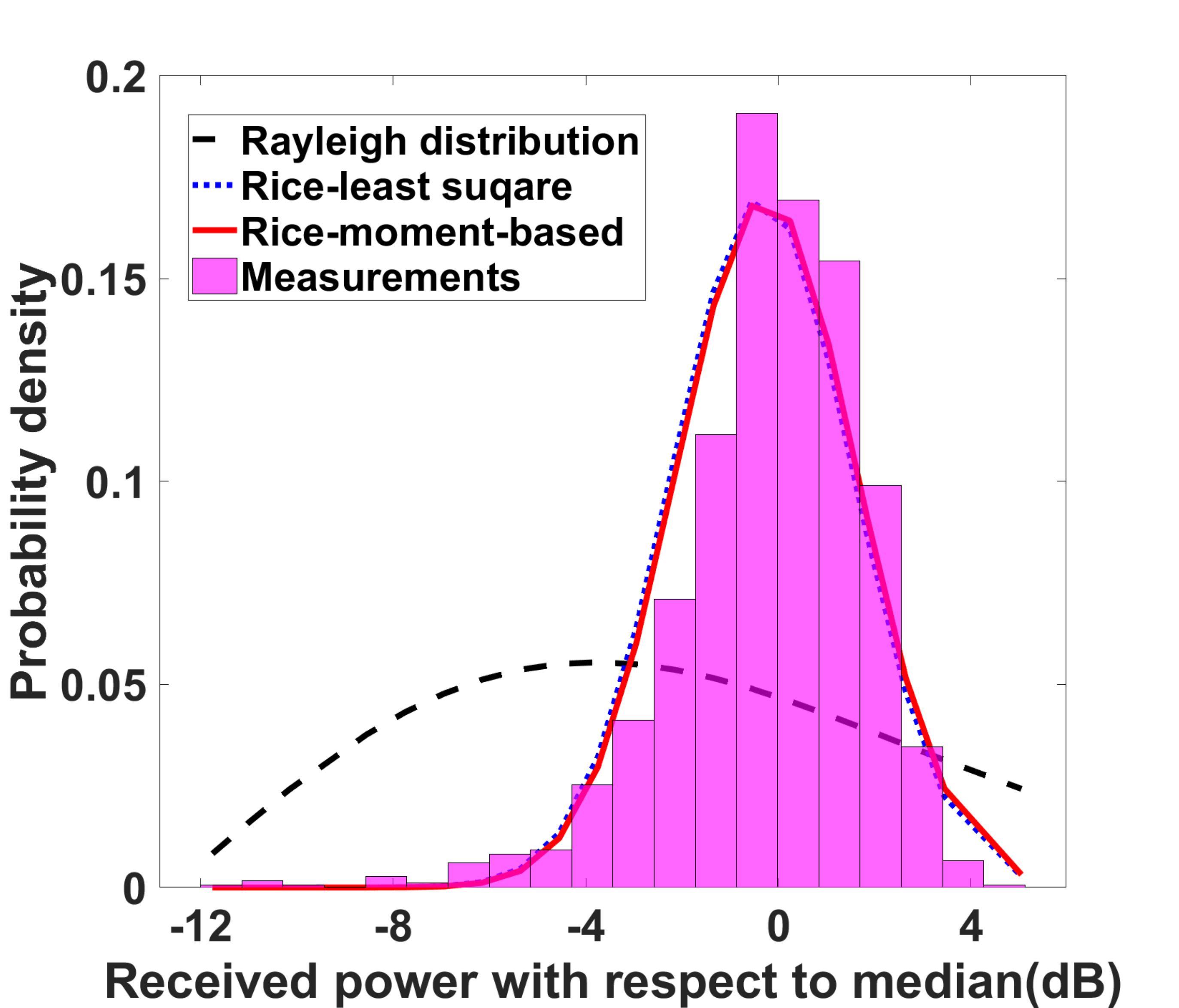}
		\label{fig:PDF2}}
	\caption{CDF and PDF of fitted Rician distribution model.}
	\label{fig:CDFandPDF}
	\vspace{-0.2in}
\end{figure*}
\begin{table}[!h]
\centering
\small
\caption{Comparison results.}
\label{tab:comparison}
\resizebox{2.6in}{!}{
\begin{tabular}{|cccc|}
\hline
LSF scenario           & Level        & \begin{tabular}[c]{@{}c@{}}$\sigma${[}dB{]}\\ General model\end{tabular} & \begin{tabular}[c]{@{}c@{}}$\sigma${[}dB{]}\\ AF model\end{tabular} \\ \hline
1(LOS)                  & 4            & -       & -       \\ \hline
2(OBS)                   & 4            & 5.17    & 4.12    \\ \hline
\multirow{3}{*}{3(NLOS)} & 3            & 4.88    & 3.76    \\  
                         & 2            & 5.7    & 4.11    \\  
                         & 1            & 6.61    & 4.32    \\ 
                         & Ground floor & 6.85    & 4.77    \\ \hline
All                      & -            & 5.84    & 4.21    \\ \hline
\end{tabular}
}
\end{table}
\linespread{1}

Based on the results above, we build a model for path loss estimation in multi-floor buildings. As the model in Eq.~\ref{eq:pathlossmodel2} is trained from office building, we test the accuracy of the model in office building. In this test, we consider two scenarios: same floor and different floor. For each scenario, the gateway is placed at a fixed position as shown in Fig.~\ref{fig:toolchain}. Then we place a LoRa node at multiple locations to record the true RSSI measurement. Meanwhile, for each location we predict the RSSI value by the model in Eq.~\ref{eq:pathlossmodel2}. The results of the same floor and different floor are shown in Fig.~\ref{fig:tool_horizontal} and Fig.~\ref{fig:tool_vertical}, respectively. We can see that the mean error in two scenarios are as low as 3.5dB and 5.4dB which indicate that the model can predict path loss in this building with high accuracy. 

\subsection{Shadowing Characteristics}
Spatial correlation characteristics of shadow fading are investigated by means of calculating the normalized autocorrelation function. To this end, one-slope models are fitted to the path-loss data of each measurement track separately. A non-fixed intercept is used in the determination of the one-slope models' parameters. As stated above, the model with non-fixed intercept provides a more accurate representation of median path loss. Then we calculate the normalized autocorrelation function $RXX(m)$ as mentioned in Sec.~\ref{subsub:shadowing}.
\begin{table}[!h]
\small
\centering
\caption{Decorrelation distance of different Buildings.}
\label{tab:decorrelation}
\begin{tabular}{|c|c|}
\hline
Building &  Decorrelation distance\\  \hline
Office                           & 2.8m        \\ \hline
Residential Building                             & 1.6m          \\ \hline
Car park                             & 1.8m         \\ \hline
Warehouse                             & 2.4m       \\ \hline
\end{tabular}
\end{table}

We now present the results of shadowing characteristics as the steps in Sec.~\ref{subsub:shadowing}. An example of the normalized autocorrelation versus distance for one measurement track is shown in Fig.~\ref{fig:autocorrelation1}. We can see a rapid decrease in autocorrelation with distance. This reinforces the popular assertion that shadow fading autocorrelation decays exponentially with distance, as reported in~\cite{jalden2007correlation,tanghe2008industrial}. Decorrelation distances, as defined in Sec.~\ref{subsub:shadowing}, are calculated for each measurement track individually. As shown in Tab.~\ref{tab:decorrelation}, decorrelation distances varied between 1.6m and 2.8m in different buildings. These are comparable with decorrelation distances in the order of 1 to 2 m reported in~\cite{jalden2007correlation} for indoor measurements at 1800 and 5200 MHz. The results indicate that the large scale fading is almost independent from one local area to another. 


\section{Temporal Fading Results}
\label{sec:temporalfadingresults}
This section discusses the temporal-fading measurements of the indoor environment. Take office building as an example, Fig.~\ref{fig:temporalfading} presents a series of typical temporal fading samples of received power measured in TF scenario 1 and 2, respectively. It can be seen in Fig.~\ref{fig:temporalfading1} that fades occur during the period of 0 to 110s, and 280s to 330s, separated by periods during which the received signal strength remained almost constant. Clearly, the variations are caused by the movement of personnel in the laboratory. The typical dynamic range for fading in TF scenario 1 is found to be about 8 dB, and can be up to 17dB occasionally. However, the temporal fading measurements in TF scenario 2 is somewhat different (Fig.~\ref{fig:temporalfading2}). Here, the received signal strength exhibits more variations, and is over a large dynamic range (typically 20 dB). This is because in TF scenario 2, there are more people moving around.

In order to verify the variations are indeed caused by people, we conduct another experiment at weekend midnight when there is no people in the office building. From the results in Fig.~\ref{fig:temporalfading3} and Fig.~\ref{fig:temporalfading4}, we can see that there is only 1-2dBm variations when there is no people in the environment. In this case, the small variations are mainly caused by environmental changes (e.g., temperature, airflow), hardware noise and thermal effects~\cite{jana2009effectiveness}.
\linespread{1.1}
\begin{table}[t]
\tiny
\centering
\caption{Rician K-factor per building.}
\label{tab:kfactor}
\begin{tabular}{|c|cc|cc|}
\hline
\multirow{2}{*}{Building} & \multicolumn{2}{c|}{Moment-based} & \multicolumn{2}{c|}{Curve fitting} \\ \cline{2-5} 
                          & TF  1           & TF 2            & TF 1             & TF 2            \\ \hline
Office Building           & 18.91           & 12.07           & 18.83            & 12.06           \\ \hline
Residential Building      & 18.62           & 23.15           & 18.64            & 23.16           \\ \hline
Car park                  & 22.64           & 11.14           & 22.65            & 11.12           \\ \hline
Warehouse                 & 17.6            & 21.43           & 17.61            & 21.44           \\ \hline
\end{tabular}
\vspace{-0.1in}
\end{table}
\linespread{1}
\begin{figure*}[!ht]
	\centering
	\subfigure[Impact of data rate]{
		\includegraphics[width=1.7in]{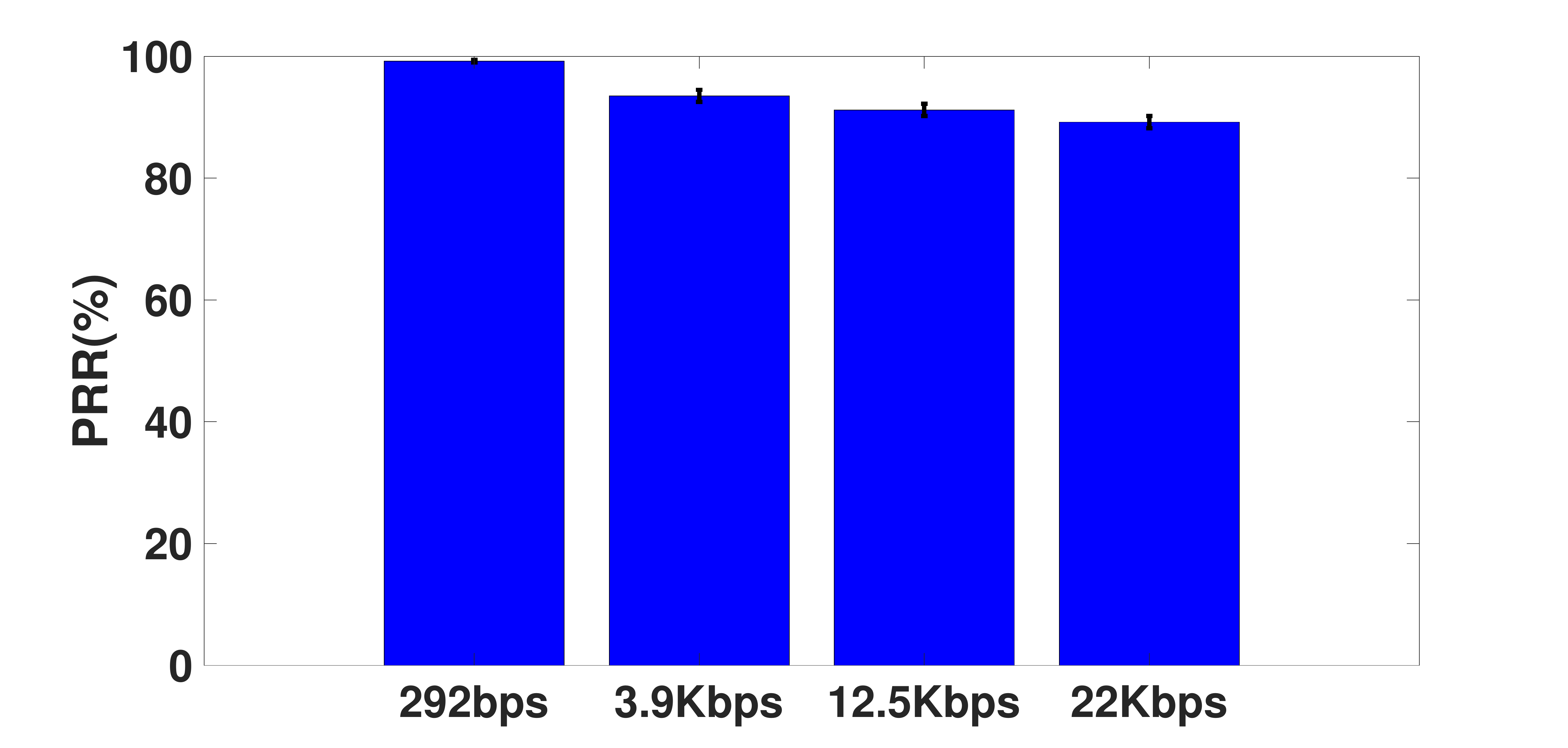}
		\label{fig:testbed_datarate}}
		\hspace{-0.2in}
		\subfigure[Impact of BW]{
		\includegraphics[width=1.7in]{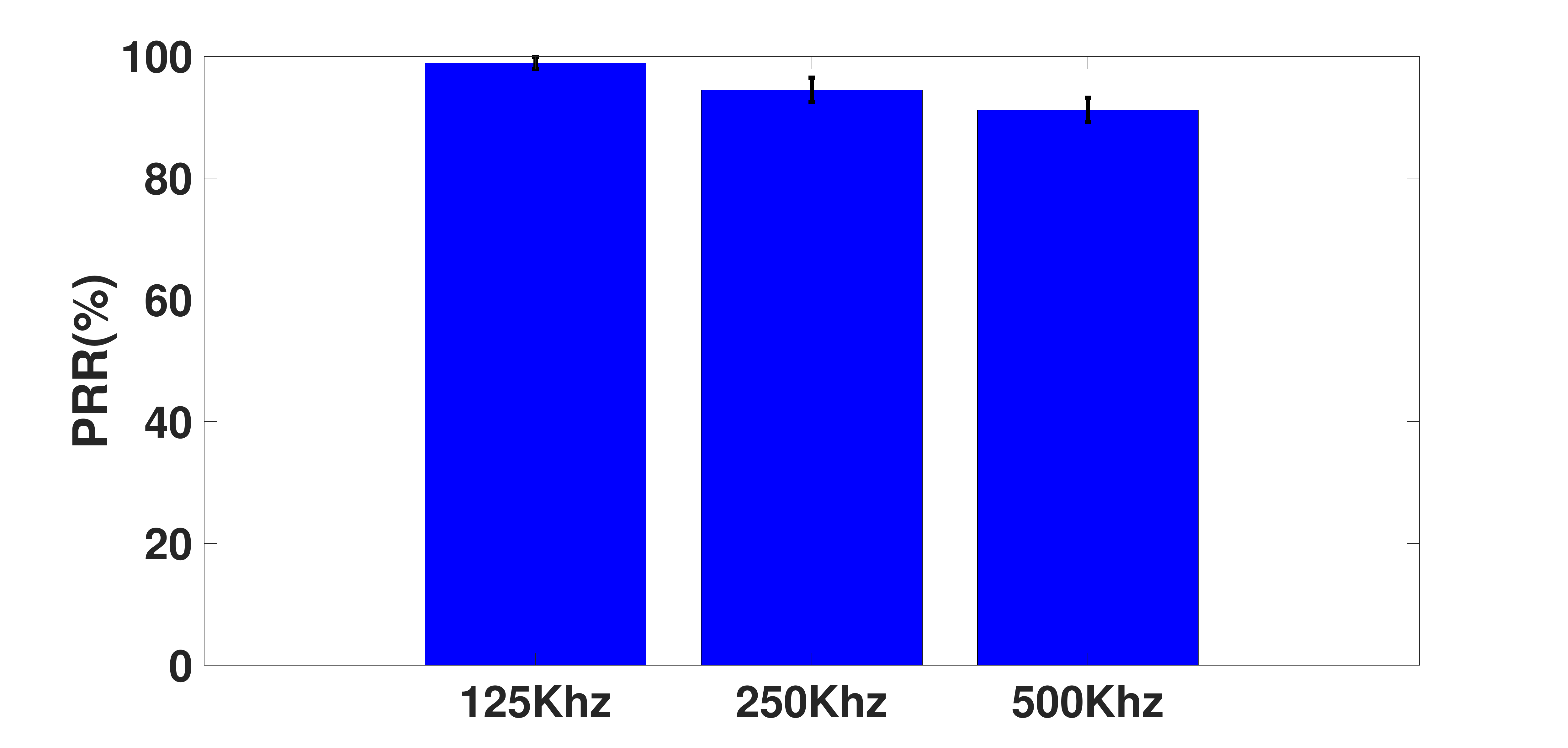}
		\label{fig:testbed_BW}}
		\hspace{-0.2in}
		\subfigure[Impact of SF]{
		\includegraphics[width=1.7in]{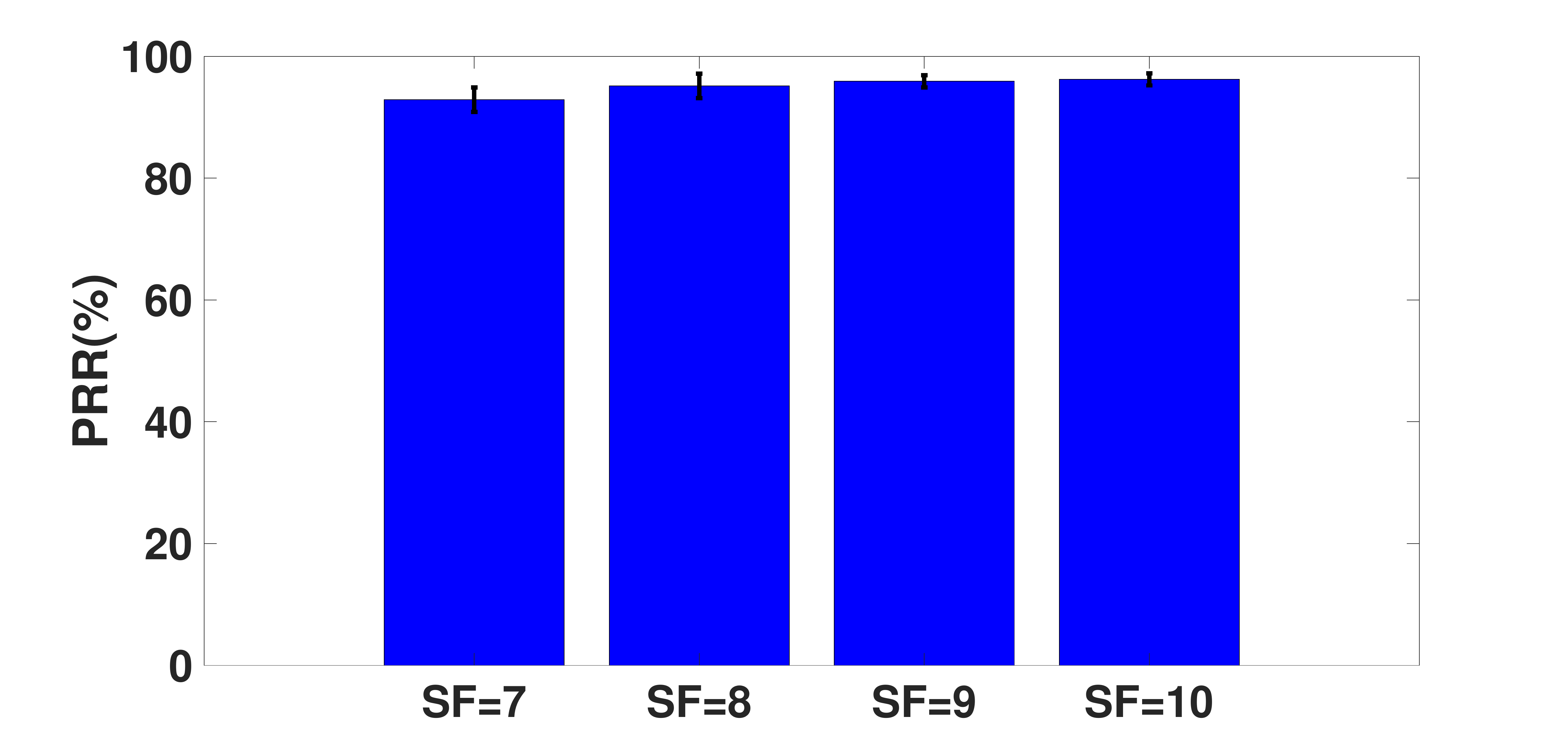}
		\label{fig:testbed_SF}}
		\hspace{-0.2in}
        \subfigure[Impact of frequency]{
		\includegraphics[width=1.7in]{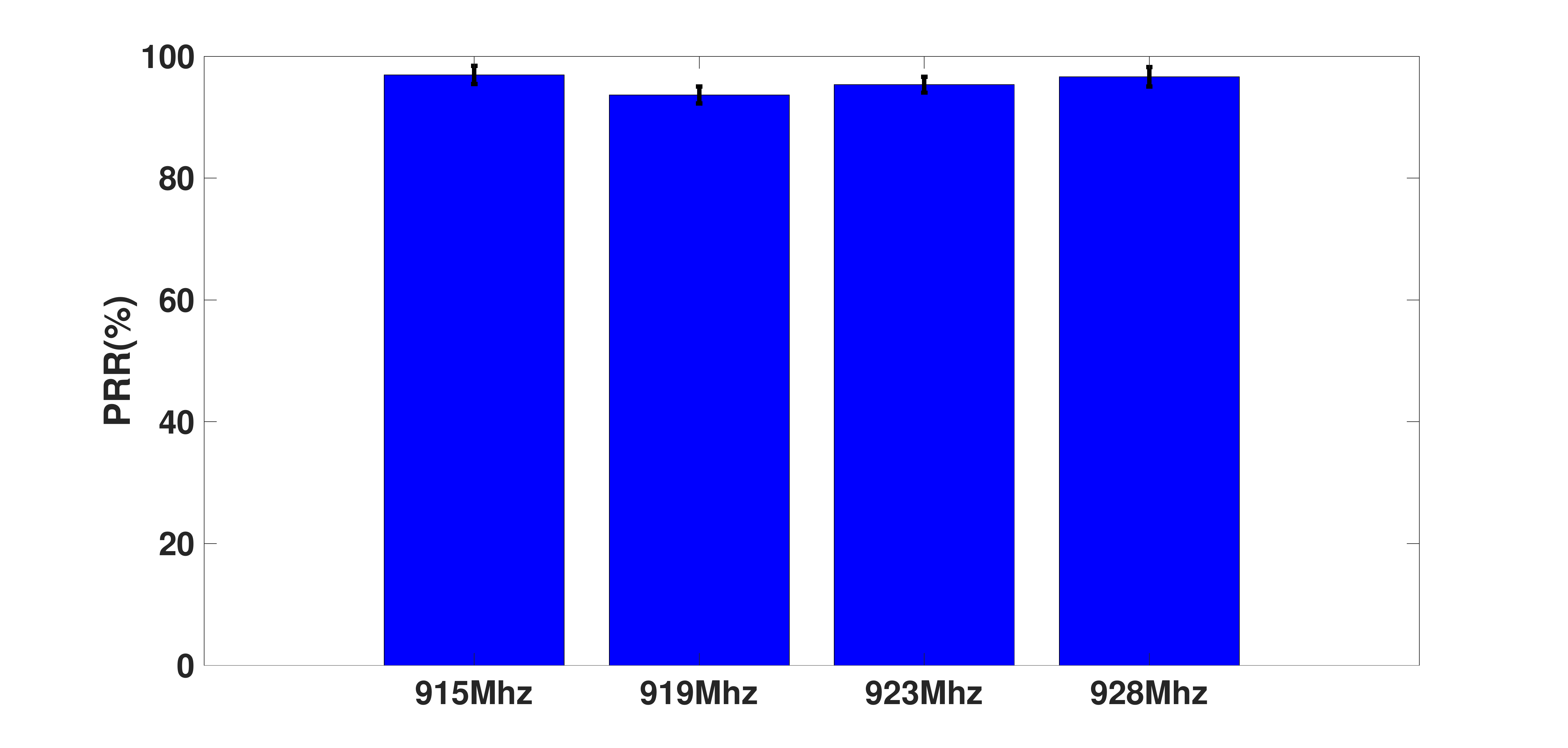}
		\label{fig:testbed_frequency}}
	\caption{Evaluation results of Testbed.}
	\label{fig:testbed}
	\vspace{-0.2in}
\end{figure*}
\begin{figure*}[!t]
	\centering
	\subfigure[Impact of data rate (Building 2)]{
		\includegraphics[width=1.8in]{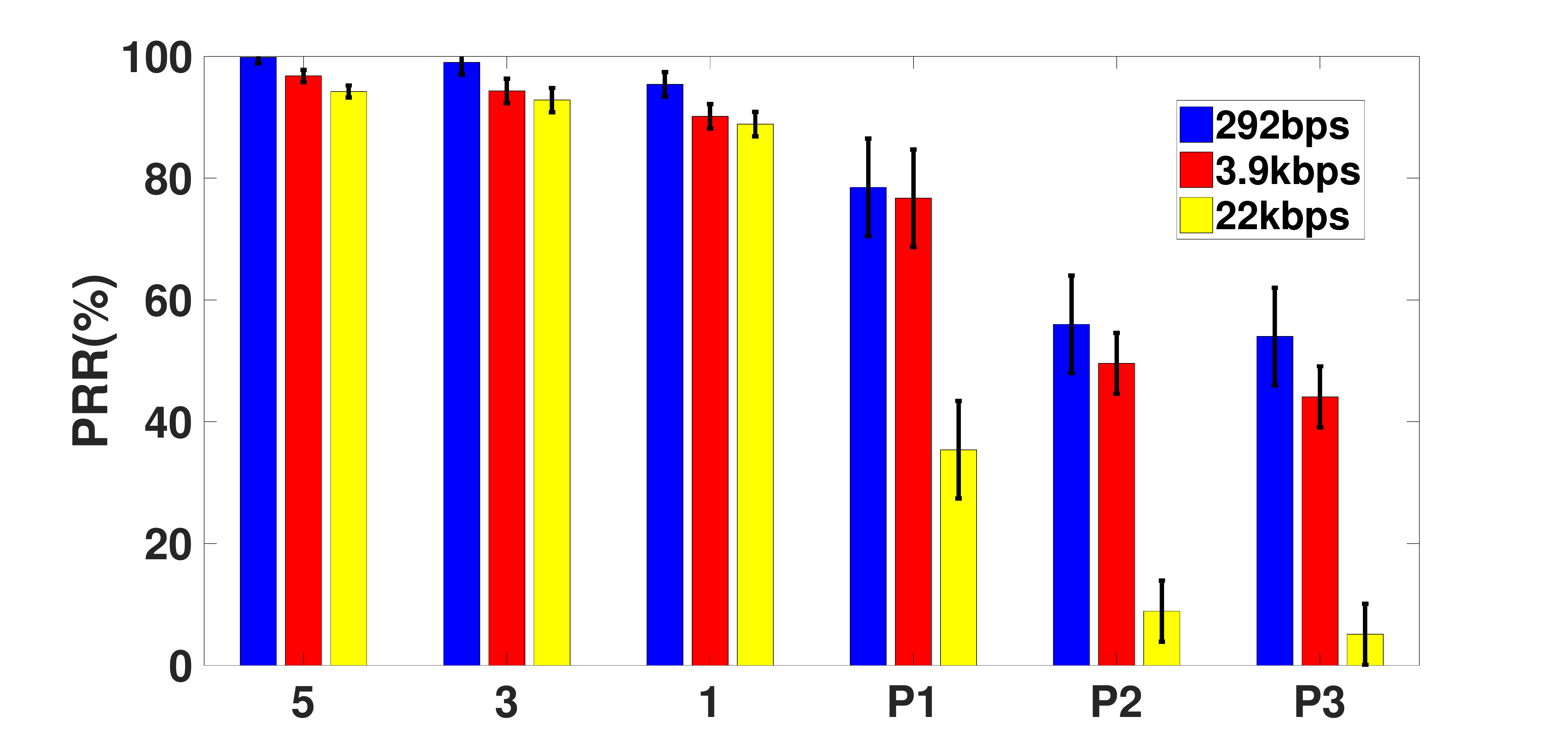}
		\label{fig:carlingford_datarate}}
		\subfigure[Impact of SF (Building 2)]{
		\includegraphics[width=1.8in]{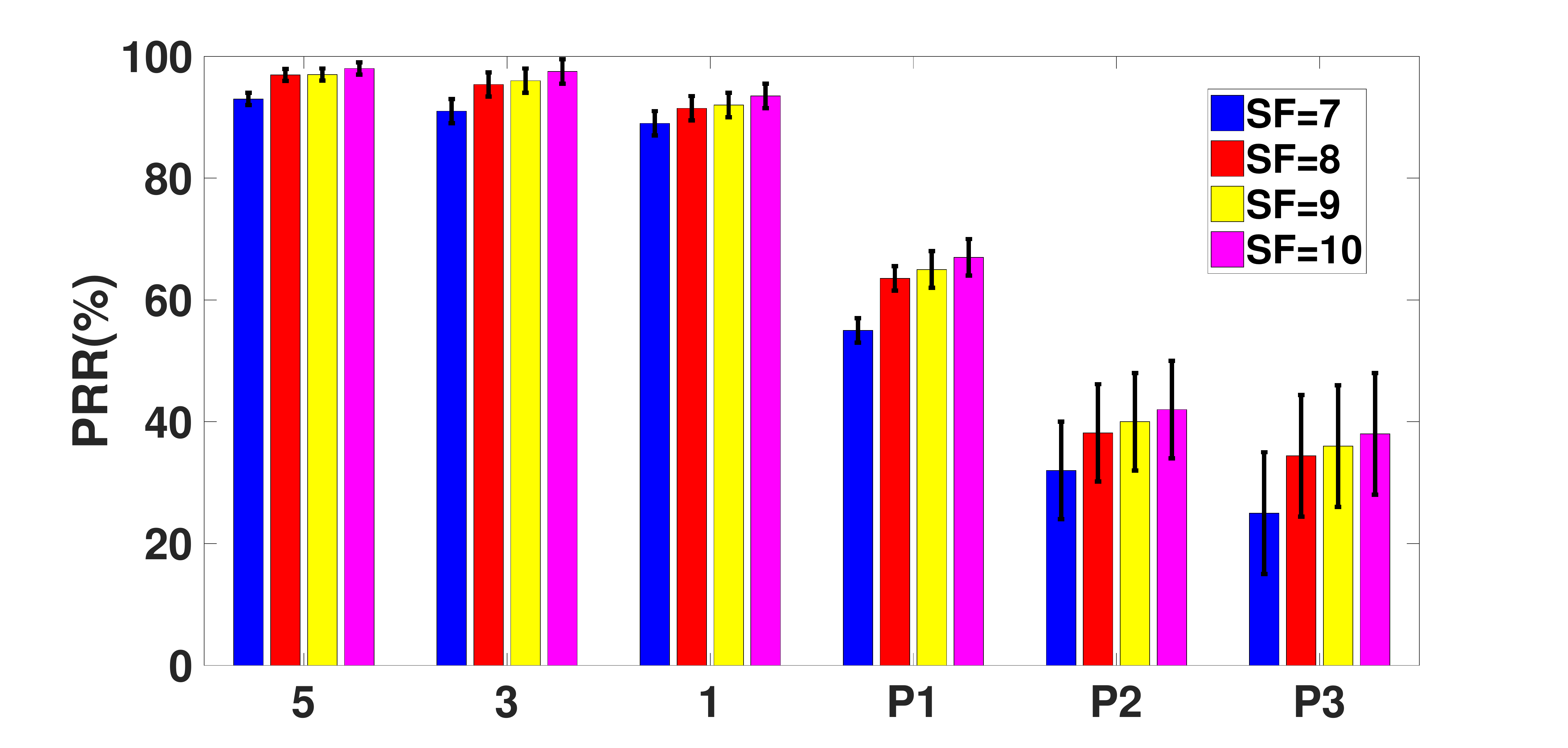}
		\label{fig:carlingford_SF}}
			\subfigure[Impact of BW (Building 2)]{
		\includegraphics[width=1.8in]{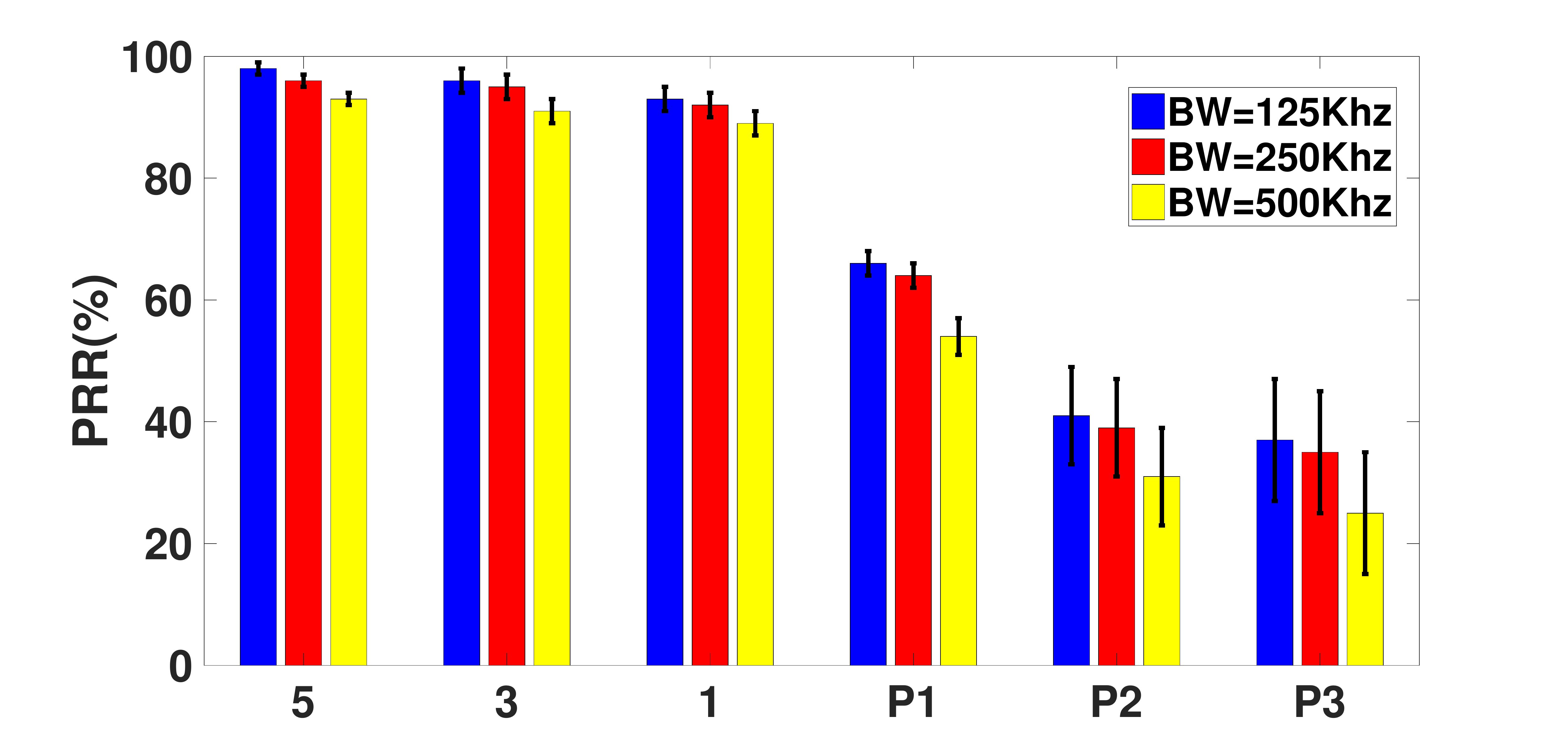}
		\label{fig:carlingford_BW}}
		\subfigure[Impact of data rate (Building 3)]{
		\includegraphics[width=1.8in]{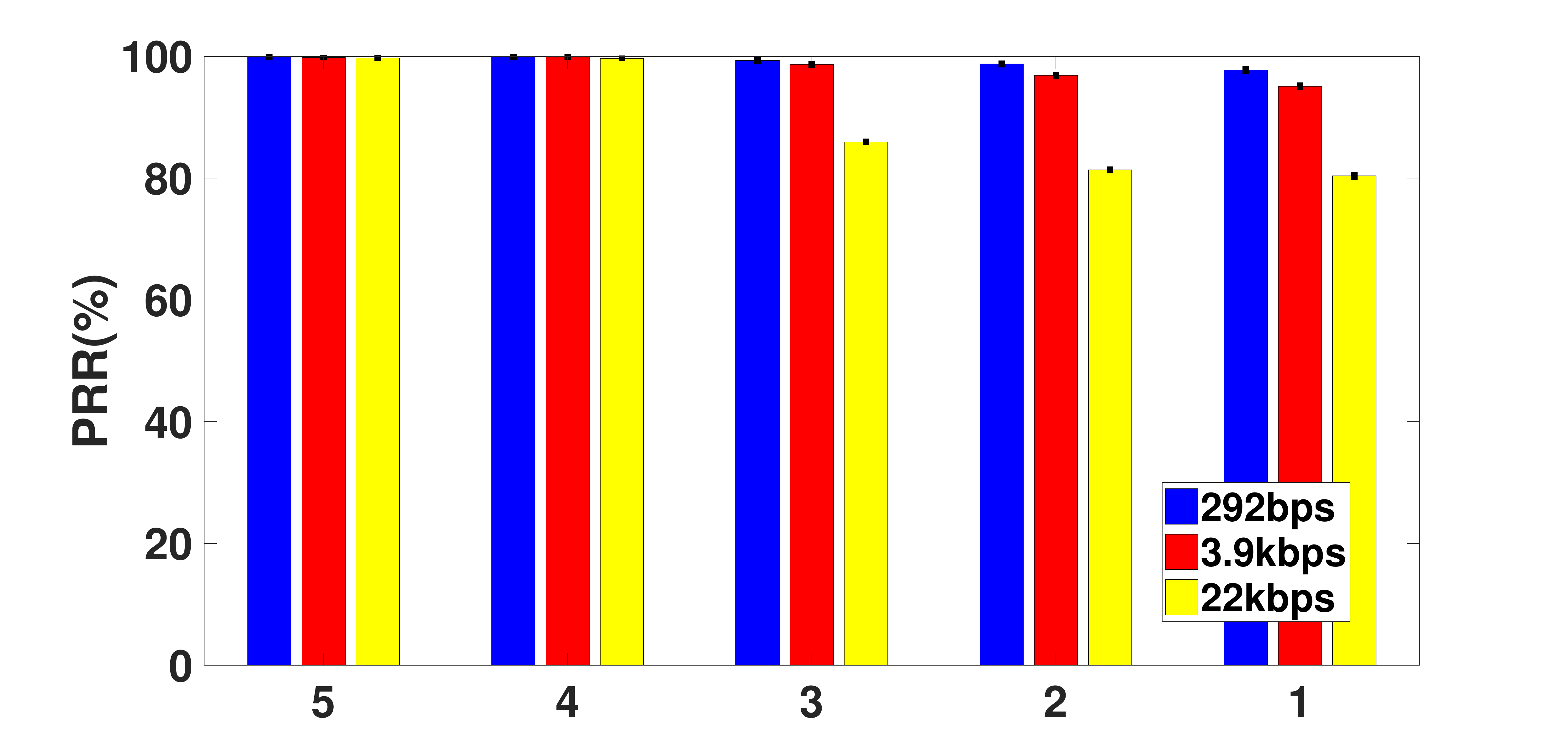}
		\label{fig:carpark_datarate}}
		\subfigure[Impact of SF (Building 3)]{
		\includegraphics[width=1.8in]{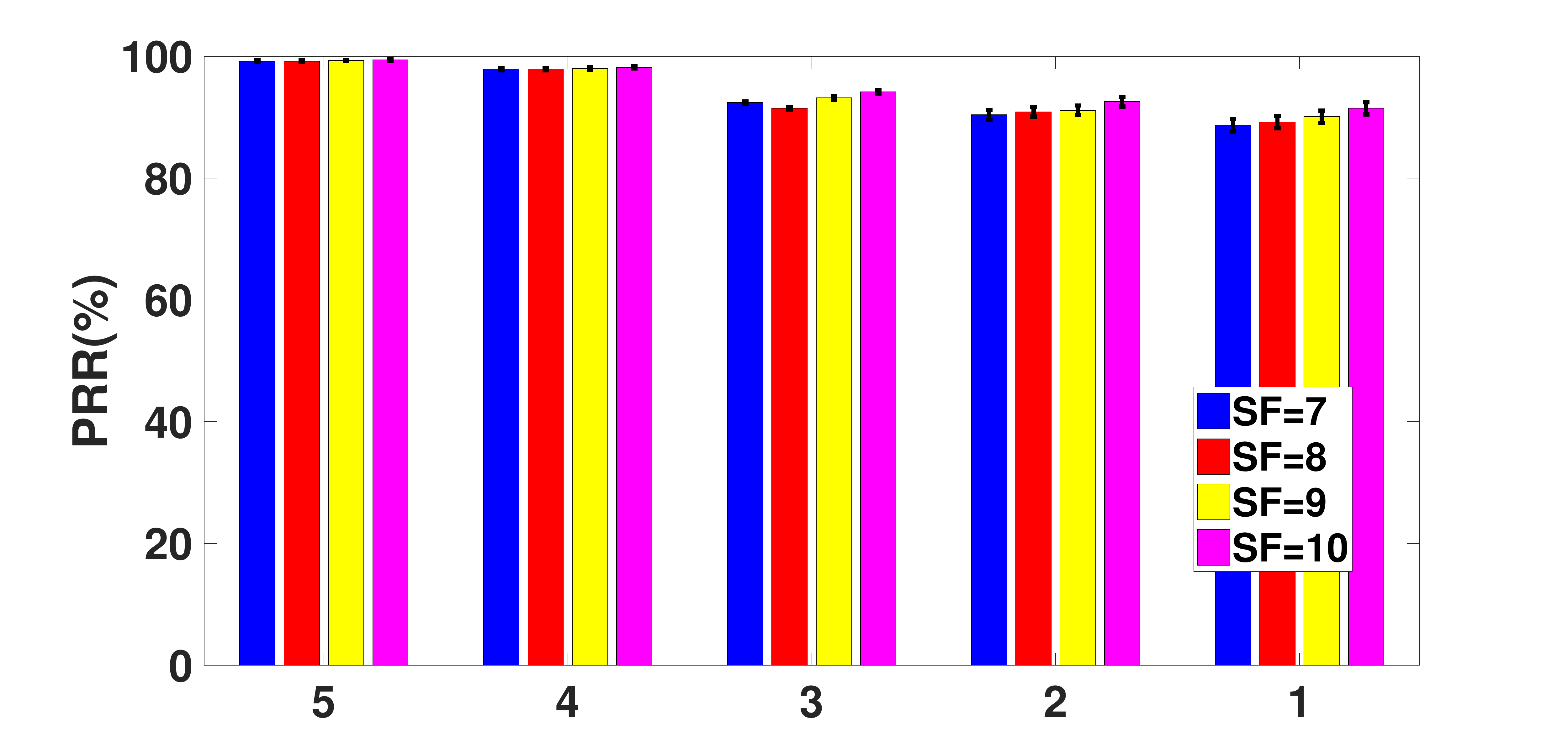}
		\label{fig:carpark_SF}}
		\subfigure[Impact of BW (Building 3)]{
		\includegraphics[width=1.8in]{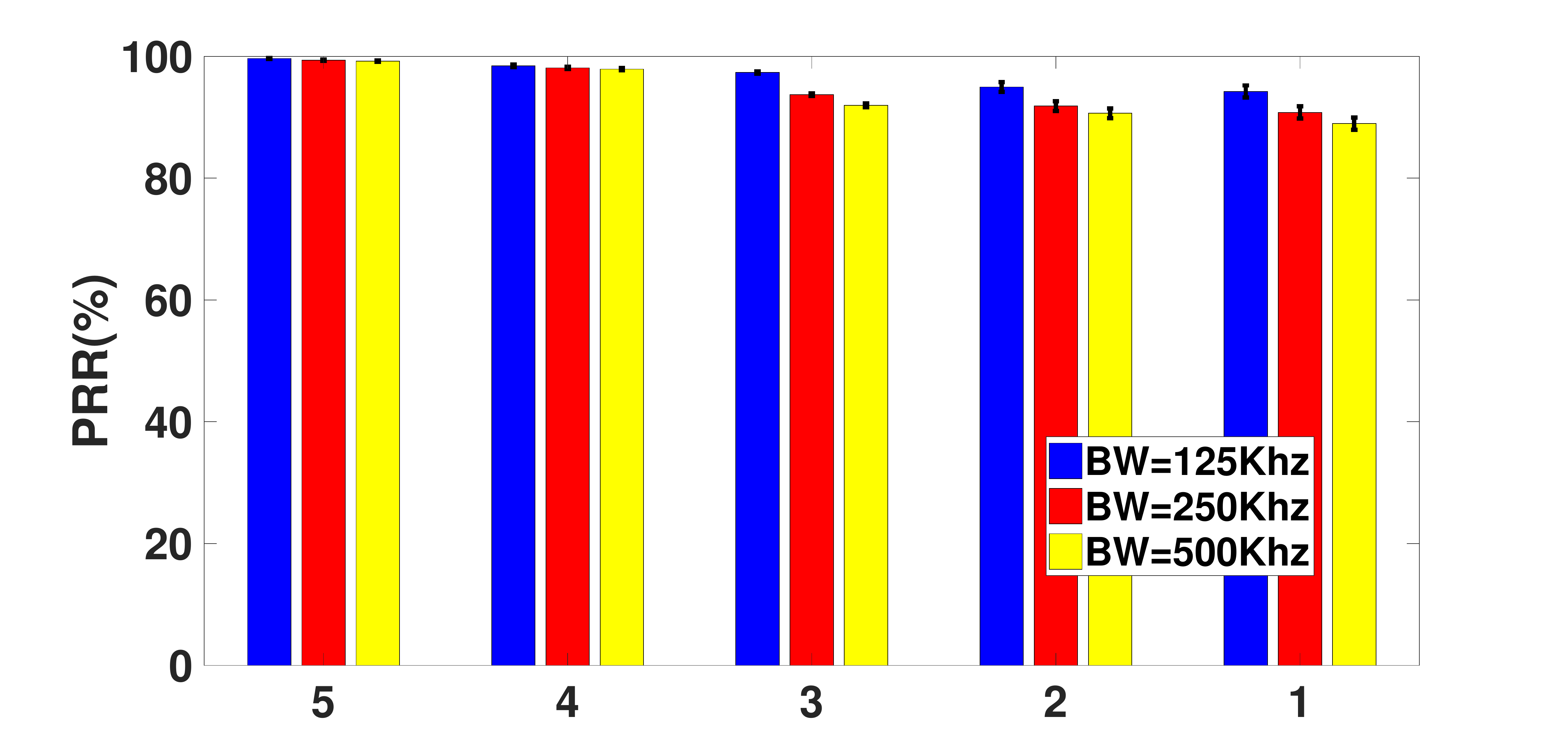}
		\label{fig:carpark_BW}}
	\caption{Evaluation results of residential building (Building 2) and car park (Building 3).}
	\label{fig:residential_carpark}
	\vspace{-0.2in}
\end{figure*}

We further calculate the Rician K-factor using the two methods mentioned early in Sec.~\ref{sub:temporal}. After obtaining the Rician K-factor, we compute the CDF and PDF of the received signal power for each scenario in different buildings. In this step, we are interested in the periods during which the fading occurs. To this end, data collected during the quiescent periods between fading variations are removed from the original samples. This is achieved by using a threshold detection algorithm to determine when the fading periods occur. 

Tab.~\ref{tab:kfactor} lists all K-factors for TF scenario 1 and 2 using the moment-based method and least-square curve fitting method. We find that these two methods obtain similar results in both scenarios. Take office building as an example, Fig.~\ref{fig:CDFandPDF} plots the CDF and PDF of the fitted Rician distribution model. We also plot the fitted Rayleigh distribution for comparison. From the good agreement between both types of estimators as well as the good correspondence between empirical and fitted Rician model in Fig.~\ref{fig:CDFandPDF}, we can draw the conclusion that the indoor environment temporal fading follows Rician distribution. Similar results are also reported in previous studies but with different K-factors~\cite{tanghe2008industrial,seidel1992914}. For example, the study in~\cite{seidel1992914} show that the Rician K-factor varies from 6dB to 12dB in a typical office environment. In an industrial environment, the K-factors are found to vary greatly from 4dB to 19dB~\cite{tanghe2008industrial}. 
\begin{table}[!hb]
\tiny
\centering
\caption{Fade margin in different buildings.}
\label{tab:fademargin}
\begin{tabular}{|c|cc|}
\hline
\multirow{2}{*}{Building} & \multicolumn{2}{c|}{Fade margin (dB)} \\ \cline{2-3} 
                          & TF  scenario 1     & TF scenario 2    \\ \hline
Office Building           & 19.4               & 17.4             \\ \hline
Residential Building      & 20.1               & 15               \\ \hline
Car park                  & 17.8               & 14.7             \\ \hline
Warehouse                 & 18.6               & 13.5             \\ \hline
\end{tabular}
\end{table}

The K-factor for the Rician model which best fits experimental results can reasonably be considered to be determined by the extent to which motion in the building alters the multi-path structure at the Rx. For example, in office building the K-factor in scenario 2 is lower than that in scenario 1. As explained in Sec.~\ref{subsub:method}, low K-factors indicates large motion in the wireless propagation environment. This is because in scenario 2, the Rx is put at ground level where the entrance of building is located. Since there are more people moving around which leads to more variations in scenario 2. Similar patterns are also observed in other buildings. The lowest K-factor occurs in TF scenario 2 of car park where Rx is located near the exit of the car park. The frequent entrance and exit of cars causes large variations in the received power.

The obtained K-factors and the corresponding CDFs are used to calculate a fade margin associated with temporal fading for a given outage probability. The outage
probability, which determines the probability that the wireless system will be out of the service (quality of service not reached) and the corresponding fade margin will be used in the link budget calculation for the network planning applications. The details of the calculation are explained in~\cite{goldsmith2005wireless}. For an outage probability of 0.01 (99$\%$ of the time, the variation around the median will not exceed the fade margin), the fade margin in different buildings is summarized in Tab.~\ref{tab:fademargin}. The fade margin can be used in link budget analysis.


\section{Coverage Experiment Results}
\label{sec:coverageresults}
\begin{figure*}[!ht]
	\centering
		\includegraphics[width=6in]{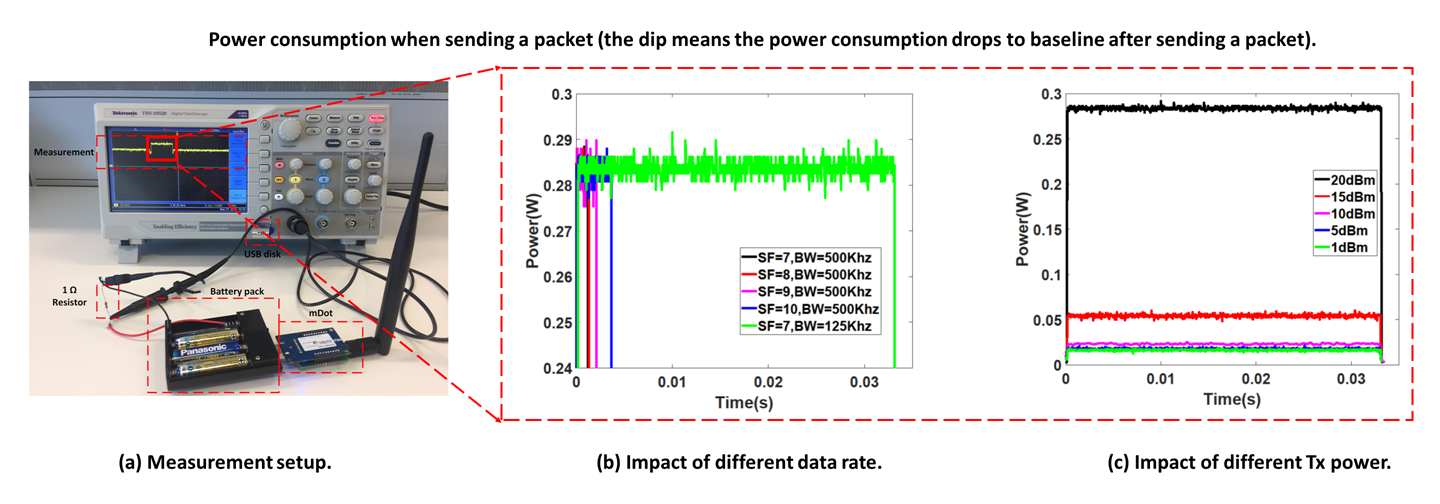}
		\vspace{-0.1in}
	\caption{Energy consumption profile.}
	\label{fig:energyconsumption}
	\vspace{-0.1in}
\end{figure*}
\linespread{1.1}
\begin{table*}[th]
\centering
\small
\caption{Energy consumption (mJ).}
\vspace{-0.1in}
\label{tab:energy}
\begin{tabular}{@{}cccccccccccc@{}}
\toprule
          & \multicolumn{2}{c}{SF=7}& & \multicolumn{2}{c}{SF=8} & & \multicolumn{2}{c}{SF=9} && \multicolumn{2}{c}{SF=10} \\ \cline{2-3} \cline{5-6} \cline{8-9} \cline{11-12} 
BW        & 125Khz     & 500Khz     & & 125Khz     & 500Khz      & & 125Khz      & 500Khz     && 125Khz      & 500Khz       \\
Data rate & 5468 bps   & 21875 bps  & & 3125 bps   & 12500 bps   & & 1757 bps    & 7031 bps   && 976 bps     & 3906 bps     \\ \cline{2-3} \cline{5-6} \cline{8-9} \cline{11-12} 
20dBm     & 9.36       & 0.33      &  & 10.22       & 0.342      &  & 10.24        & 0.57        && 11.98        & 1.02        \\
15dBm     & 1.29        & 0.018     &     & 1.47       & 0.0294   &       & 1.66         & 0.059         && 1.91         & 0.105           \\
10dBm     & 0.84        & 0.0059    &       & 0.86        & 0.0116 &        & 0.91         & 0.024        & & 0.96        & 0.0421           \\
5dBm      & 0.55        & 0.0055    &       & 0.57        & 0.0114 &        & 0.61         & 0.0204       & & 0.65         & 0.0407           \\
1dBm      & 0.43        & 0.0053    &     & 0.47        & 0.0108   &      & 0.53         & 0.0195        && 0.54         & 0.0341          \\ \bottomrule
\end{tabular}
\vspace{-0.2in}
\end{table*}
\linespread{1}
\subsection{Coverage Analysis}
\subsubsection{Results of Office Building}
Fig.~\ref{fig:testbed} plots the evaluation results with average values and $95\%$ confidence level. In the following, we analyze the impact of data rate, spread factor (SF), bandwidth (BW), and frequency in turn.

Fig.~\ref{fig:testbed_datarate} shows the PRR using different data rate. We notice that the higher the data rate is, the lower the PRR is. For example, the data rate of 292bps achieves the best PRR while the highest data rate (i.e., 22kbps) has the lowest PRR. This results correspond to the LoRa characteristics: a lower data rate  has stronger penetration ability and can achieve a longer communication distance. Followed by that, Fig.~\ref{fig:testbed_BW} plots the impact of BW on PRR. We find that the greater the bandwidth is, the lower PRR it achieves. This result corresponds well with the results in~\cite{petajajarvi2017evaluation}. This is because the symbol rate $R_{s}=\frac{BW}{2^{SF}}$; therefore, when the transmission power is constant, the wider the bandwidth is, the less energy will be distributed for each symbol. From Fig.~\ref{fig:testbed_SF}, we find the PRR increases slightly when SF increases from 7 to 10. The reason is the same: when the transmission power is constant, the larger the SF is, the more energy will be distributed for each symbol.

The PRR of different center frequencies are plotted in Fig.~\ref{fig:testbed_frequency}. It can bee seen that the PRR of channel 919Mhz is slightly lower than the other channels. This may be caused by the following reasons: it can be due to the different antenna efficiency or amplification gains for the frequencies, due to the interference from the environment, or due to the differences in the radio frequency propagation for this frequency.
\subsubsection{Results of Residential Building and Car Park}
The results of residential building and car park are shown in Fig.~\ref{fig:residential_carpark}. We can see that the impact of data rate, SF and BW in these two buildings are the same as observed in office building. For example, the data rate of 292bps consistently achieves the best PRR across all the floors while the highest data rate (i.e., 22kbps) has the lowest PRR in these two buildings. Another finding is that in the residential building, the PRR in basement P1 P2 and P3 are significantly lower than that of levels above ground which suggests underground is indeed a challenging communication environment.
   
\subsection{Energy Consumption Analysis}
In this subsection, we analyse the energy consumption of different communication settings. The experiment setup for the power measurement is shown in Fig.~\ref{fig:energyconsumption}. In order to capture both the average current and the time requirement for each transmission event, the Tektronix TBS1052B oscilloscope is used. As shown in the figure, we connect the mDot with a 1$\Omega$ resistor in series and power it using a 4.5V AA battery pack. The oscilloscope probe is then connected across the resistor to measure the current going through.  The node keeps sending a 16 byte packet every 1s using different SF, BW and transmit power. The power consumption profile is stored in a USB disk for further analysis.

The SF and BW determine the data rate which further determines on-air time and consequentially the amount of the energy consumed to send a packet. 
Tab.~\ref{tab:energy} summarizes the energy consumption under different parameter settings. From the results, we have the following two findings. First, when the node is transmitting at a constant data rate, there is a significant gap between the maximum transmitting power and others. For example, when using 976bps (SF=10, BW=125Khz) it consumes 11.98mJ to transmit a 16 byte packet while it only consumes 1.91mJ using 15dBm transmitting power which means it consumes 6$\times$ less energy. Second, when the transmission power is constant, the energy consumption using the minimum data rate is remarkably higher than using the maximum data rate. For example, when using 10dBm transmitting power, the energy consumption using the minimum data rate (SF=10, BW=125Khz) is 0.86mJ which is 145$\times$ higher than that using the maximum data rate (SF=7, BW=500Khz) which is only 0.0059mJ. 

To provide an easier interpretation of the results, we plot the energy consumption profile in Fig.~\ref{fig:energyconsumption}. We can see that the transmission time significantly increases with the decrease of data rate. Therefore, low data rate will lead to longer on-air time and consequently consumes more energy. Also, it can be seen that the amount of energy consumed for sending the same packet using the minimum and the maximum transmit power differs by 17$\times$. The results emphasizes the significance of choosing an appropriate transmission power and enabling the LoRa adaptive data rate feature in the energy-constrained applications.

\section{Related work}
\label{related}

As LoRaWAN aims at long distance wide area network, researchers have conducted extensive tests in wide area outdoor environment to understand the performance and limitation of LoRaWAN. For example, ~\cite{augustin2016study} provides a comprehensive understanding of the LoRa modulation, including the data rate, frame format, spreading factor, receiver sensitivity, etc. The researchers in~\cite{bor2016lora} study the range of reliable links, the receiver sensitivity as well as LoRa scalability. In a similar work~\cite{blenn2017lorawan}, the researchers analyze packet payloads, radio-signal quality, and spatiotemporal aspects, to model and estimate the performance of LoRaWAN. Researchers also studied the path loss model of LoRa in outdoor environment~\cite{linka2018path,petajajarvi2015coverage}. Different from these works, our work focuses on the path loss model in indoor environment. Similar to our study, Lukas et al. also measure the indoor signal propagation characteristics of LoRa Technology in a building~\cite{gregora2016indoor}. However, their measurement is relatively simple and can just provide a rough understanding of signal propagation in a building. In another similar work, Said et al. investigate path loss and temporal fading of LoRa mote in dairy barns~\cite{benaissa2017internet}.

Pierre et al. study the performance of LoRaWAN unconfirmed uplink data frames in an indoor environment~\cite{neumann2016indoor}. In another work~\cite{cattani2017experimental}, the authors focus their evaluation on the impact of physical layer settings on the effective data rate and energy efficiency of communications. Simulation is a useful tool to understand the performance of wireless network as field test takes too much effort and a large testbed in real environment is not always available. For instance, the authors in~\cite{adelantado2017understanding} investigate the limitation of LoRaWAN via simulation and point out several open research challenges. Guillaume studied the collision and packet loss of LoRaWAN network via theoretical analysis~\cite{ferre2017collision}. Because LoRa technology is patented, only a few details about its operations are actually available. Many researchers have tried to reverse-engineering the technology using Software Defined Ratio (SDR) and they have successfully used SDR to encode/decode LoRa signal~\cite{decodinglora,LoraSDR}.


\section{Conclusion}
\label{sec:conclusion}
Understanding the performance and characterization of LoRa technology in indoor buildings is imperative for its deployment and application. In this paper, we have presented a comprehensive and sophisticated study regarding the large scale fading characteristics, temporal fading characteristics, and coverage of LoRa technology in four types of multi-floor buildings, as well as energy consumption using different communication modes. The results are originally used to help our industry partner design and optimize their LoRa-based smart building network. We hope the findings presented in this paper can also provide insights into the development of practical LoRa-based indoor applications. 

\tiny{
\bibliographystyle{IEEEtran}
\bibliography{draft}
}
\end{document}